\documentclass[a4paper,12pt]{article}
\pdfoutput=1
\usepackage{graphicx,subfigure,amsmath,amssymb,multirow}
\usepackage{cite}

\newlength{\dinwidth}
\newlength{\dinmargin}
\def\be{\begin{equation}}
\def\ee{\end{equation}}
\def\ba{\begin{eqnarray}}
\def\ea{\end{eqnarray}}
 \def\la{ \langle}
  \def\ra{ \rangle}
     \def\e{ \epsilon}
      \def\r{ \gamma}
       \def\lbd{\lambda}
        \def \d {{\rm d}}
           \def\w{\omega} 
            \def\u{\mu}
              \def\a{\alpha}
  \def\b{\beta}
\def\v{\nu}
     \def\ve{ \varepsilon}  

\usepackage{color}

\allowdisplaybreaks

\begin{document}
\title{\bf  Revisiting the form factors of $P\to V$ transition within the light-front quark models}
\author{Qin Chang$^{a,b}$\footnote{changqin@htu.edu.cn}, Xiao-Nan Li$^{a}$ and Li-Ting Wang$^{a}$\\
{ $^a$\small Institute of Particle and Nuclear Physics, Henan Normal University, Henan 453007,  China}\\
{ $^b$\small Institute of Particle Physics and Key Laboratory of Quark and Lepton Physics~(MOE) }\\[-0.2cm]
{ \small Central China Normal University, Wuhan, Hubei 430079, China}}
\date{}

\maketitle
\begin{abstract}
We  investigate the self-consistency and Lorentz covariance of the covariant light-front quark model~(CLF QM) via the matrix elements and form factors~(${\cal F}=g$, $a_{\pm}$ and $f$) of $P\to V$ transition. Two types of correspondence schemes between  the  manifest covariant Bethe-Salpeter  approach and the light-front quark model are studied. We find that, for $a_{-}(q^2)$ and $f(q^2)$,  the CLF results obtained via $\lbd=0$ and $\pm$ polarization states of vector meson within the traditional type-I correspondence scheme  are inconsistent with each other; and moreover, the strict covariance of the matrix element is violated due to the nonvanishing spurious contributions associated with noncovariance. We further show that such two problems have the same origin and can be resolved simultaneously by employing the type-II correspondence scheme, which advocates an additional replacement $M\to M_0$ relative to the traditional  type-I scheme; meanwhile, the results of  ${\cal F}(q^2)$ in the standard light-front quark model~(SLF QM) are exactly the same as the valence contributions  and equal to numerally   the full results in the CLF QM, {\it i.e.}, $[{\cal{F}}]_{\rm SLF}=[{\cal{F}}]_{\rm val.}\doteq[{\cal{F}}]_{\rm full}$. The numerical results for some $P\to V$ transitions are updated within the type-II scheme. Above findings confirm  the conclusion obtained via the decay constants of vector and  axial-vector  mesons in the previous works.

\end{abstract}

\newpage
\section{Introduction }
The form factor and decay constant are important physical quantities in understanding the internal structure of hadrons, and play crucial roles for predicting the observables of meson decays.  It is well-known that they must be treated with a nonperturbative method. There are many different candidates for this purpose, such as Wirbel-Stech-Bauer model~\cite{Wirbel:1985ji}, lattice calculations~\cite{Daniel:1990ah}, vector meson dominance model~\cite{Ametller:1993we,Gao:1999qn}, perturbative QCD  with some nonperturbative inputs~\cite{Lepage:1980fj,Li:1992nu}, QCD sum rules~\cite{Shifman:1978bx,Shifman:1978by} and light-front quark models~(LF QMs)~\cite{Terentev:1976jk,Berestetsky:1977zk,Cheng:1997au,Carbonell:1998rj,Jaus:1999zv}. The traditional LF QM, {\it i.e.}, the so-called standard light-front quark model~(SLF QM), proposed by Terentev and Berestetsky~\cite{Terentev:1976jk,Berestetsky:1977zk} is a relativistic quark model based on the LF formalism~\cite{Dirac:1949cp} and LF quantization of QCD~\cite{Brodsky:1997de}. It provides a conceptually simple and phenomenologically feasible framework for the determination of  form factor, decay constant and distribution amplitude {\it et al.}, which are further applied to phenomenological researches~\cite{Jaus:1989au,Jaus:1989av,Jaus:1991cy,Jaus:1996np,Cheng:1996if,ODonnell:1996sya,Cheung:1996qt,Choi:1996mq,Choi:1997iq,Choi:1998jd,DeWitt:2003rs,Choi:2007yu,Choi:2007se,Barik:1997qq,Hwang:2000ez,Hwang:2010hw,Hwang:2001zd,Hwang:2009cu,Geng:2001de,Chang:2016ouf,Chang:2017sdl,Chang:2018aut,Chang:2018mva,Wang:2017mqp}.  However, in the SLF QM, the Lorentz covariance of the matrix element is lost since it contains a spurious dependence on the orientation of the light-front~(LF) defined in term of the light-like four-vector~$\w$ by $\w\cdot x =0$, and moreover the zero-mode contributions can not be determined. 

In order to treat the complete Lorentz structure of a matrix element and evaluate the zero-mode contributions, many efforts have been made in the past years~\cite{Cheng:1997au,Carbonell:1998rj,Jaus:1999zv,Karmanov:1991fv,Karmanov:1994ck,Karmanov:1996un,Choi:1998nf}. In Ref.~\cite{Carbonell:1998rj}, Carbonell, Desplanques, Karmanov and Mathiot~(CDKM) have developed a method based on the covariant  LF framework to identify and separate the spurious contributions and to determine the $\w$-independent physical contributions, while the zero-mode contributions  are not fully considered still. In Ref.~\cite{Jaus:1999zv}, a basically different technique is developed by Jaus to deal with the covariance and zero-mode problems with the help of a manifestly covariant Bethe-Saltpeter approach as a guide to the calculation.  In the Jaus' prescription for the covariant light-front quark model~(CLF QM), the zero-mode contributions can be well determined, and the result of the matrix element is expected to be covariant because  the spurious contribution proportional $\w$ can be eliminated  by the inclusion of zero-mode contributions~\cite{Jaus:1999zv}. This CLF QM has been used extensively to study the weak and radiative decays, as well as the other features, of hadrons~\cite{Bakker:2000pk,Bakker:2002mt,Bakker:2003up,Choi:2004ww,Choi:2009ai,Choi:2009ym,Choi:2010ha,Choi:2010zb,Choi:2011xm,Choi:2010be,Choi:2014ifm,Choi:2017uos,Choi:2017zxn,Ryu:2018egt,Hwang:2001hj,Hwang:2001wu,Hwang:2010iq,Cheung:2014cka,Wang:2008ci,Wang:2008xt,Shen:2008zzb,Wang:2009mi,Cheng:2017pcq,Kang:2018jzg,Verma:2011yw,Shi:2016gqt,Wang:2018duy,Jaus:2002sv}.  

However, it has been noted that  there still exist some problems about the self-consistency~\cite{Cheng:2003sm,Choi:2013mda,Chang:2018zjq} and strict covariance \cite{Jaus:1999zv,Chang:2018zjq} in the CLF QM. In Ref.~\cite{Cheng:2003sm}, the authors have found that the CLF results for the vector~($V$) meson decay constant obtained receptively via the longitudinal ($\lbd=0$) and the transverse ($\lbd=\pm$) polarization state are  inconsistent with each other,  $[f_V]_{\rm CLF}^{\lbd=0}\neq [f_V]_{\rm CLF}^{\lbd=\pm}$,  because the former receives an additional contribution characterized by the $B_1^{(2)}$ function~(the $B$ functions are given by Eq.~\eqref{eq:ABfuns} ), which provides about $10\%$ correction.  This inconsistency problem exists not only in the vector system but also in the axial-vector~(A) system~\cite{Chang:2018zjq}. Besides, the strict Lorentz covariance is another challenge to the  CLF QM~\cite{Jaus:1999zv,Chang:2018zjq}.  A known example is the matrix element, ${\cal A}_V^\u \equiv \la 0 | \bar{q}_2 \r^\u q_1 |V\ra$\cite{Chang:2018zjq}. Although the main $\w$ dependences are associated with the $C$ functions and can be eliminated by the zero-mode contributions~\cite{Jaus:1999zv}, there are still some residual $\w$ dependences due to the nonvanishing spurious contributions associated with  $B_1^{(2)}$ function, which violate the covariance of CLF result for ${\cal A}_V^\u$~\cite{Jaus:1999zv,Chang:2018zjq}. In order to resolve these problems, some efforts have been made.

 In the CLF QM, a manifestly covariant Bethe-Salpeter~(BS) approach is used to guide the corresponding light-front calculation,  but still using the same vertex functions and operators as employed in the SLF QM~\cite{Jaus:1999zv}. 
  Taking the vector meson as an example, the correspondence scheme~(type-I)~\cite{Jaus:1999zv, Cheng:2003sm,Choi:2013mda}
\begin{align} \label{eq:type-I}
\sqrt{2N_c}\frac{\chi_V(x,k_{\bot})}{1-x} \;\to\;  \frac{ \psi_V(x,k_{\bot})}{\sqrt{x(1-x)} \hat{M}_0}\,,\qquad D_{V,{\rm con}}  \;\to\; D_{V,{\rm LF}}\,,\qquad (\text{type-I})
\end{align}
between the  covariant BS  model and the LF QM is used in the traditional CLF QM, where the factors $D_{V,{\rm con}}=M+m_1+m_2$ and $D_{V,{\rm LF}}=M_0+m_1+m_2$ appear in the vertex operator. Within this type-I correspondence scheme, the CLF result for $f_V$ suffers from  above-mentioned self-consistency and covariance problems~\cite{Cheng:2003sm,Choi:2013mda}.   It should be noted that a significant difference between the covariant BS approach and the LF QM is that the constituent quarks of a bound-state are allowed to be off mass-shell in the former, but are required be on their respective mass-shell in the latter. Therefore, a generalized correspondence scheme~(type-II)
\begin{align} \label{eq:type-II}
\sqrt{2N_c}\frac{\chi_V(x,k_{\bot})}{1-x} \;\to\;  \frac{ \psi_V(x,k_{\bot})}{\sqrt{x(1-x)} \hat{M}_0}\,,\qquad M  \;\to\; M_0\,.\qquad (\text{type-II})
\end{align}
  is suggested by Choi and Ji~\cite{Choi:2013mda}. It is interesting that this new correspondence scheme provides a solution to the self-consistency problem of  $f_V$~\cite{Choi:2013mda}.  

In our previous work~\cite{Chang:2018zjq}, we have studied the self-consistency and the strict covariance simultaneously via the decay constants of pseudoscalar, vector and  axial-vector  mesons. It is found that~\cite{Chang:2018zjq}: (i) the  problem of self-consistency  exists not only in the vector system but also in the axial-vector system when the type-I correspondence scheme is used, but both of them can be resolved by employing the type-II scheme, which confirms Choi's findings~\cite{Choi:2013mda};  the replacement $M\to M_0$  in the  type-II scheme plays a crucial role in resolving these problems. 
 (ii)  The violation of the manifest covariance of the CLF QM with type-I scheme is caused by the same reason  as for the self-consistency problem, and the strict covariance can be recovered by taking the type-II correspondence. 
 (iii)  In addition, a clear relation between the SLF and CLF results are found
\begin{align}
[{\cal Q}]_{\rm SLF}=[{\cal Q}]_{\rm val.}\doteq[{\cal Q}]_{\rm full},\label{eq:sc}
\end{align}
within the type-II scheme, where ${\cal Q}=f_{V,A}$; 
 the subscripts ``full'' and ``val.'' denote the full result and the valence contribution in the CLF QM, respectively; and the symbol ``$\doteq$''  denote that the two quantities are equal to each other only numerically but not formally.

Besides decay constant, the form factor is another important quantity for testing the  performance of the LF QMs.  The $P\to V$ transition is also related to the spin-1 system, therefore it is worth to test  whether the form factors of  $P\to V$ transition have the problems of self-consistency and covariance, as the case of $f_{V,A}$ mentioned above, in the CLF QM with type-I correspondence scheme, and whether the type-II scheme can give a solution to these problems still. Moreover, the form factors of $P\to V$ transition are related to not only $B_1^{(2)}$ but also $B_{3}^{(3)}$ function, in which only the former contributes to  $f_{V,A}$. Therefore, the form factors of $P\to V$ transition may present much stricter test on the self-consistency and covariance of CLF QM, as well as above-mentioned  findings obtained via $f_{V,A}$. In addition, it is claimed in Ref.~\cite{Jaus:1999zv} that the form factor $a^{P\to V}_-(q^2)$ is impossible to be calculated  in the SLF formalism, which need to be checked.  In this paper, these issues will be studied in detail. 

Our paper is organized as follows. In section 2, we would like to review briefly the SLF and the CLF QMs for convenience of discussion. In section 3, the SLF and CLF results, as well as the valence contributions, are presented;  after that, the self-consistency and covariance of CLF results for the form factors of $P\to V$ transition are discussed in detail. Finally, our conclusions are made in section 4. 

\section{Brief review of light-front quark models}
\label{sec:2}
In this section, we would like  review briefly the LF QMs for calculating the current matrix element defined as 
\begin{align}\label{eq:amp1}
{\cal B} \equiv \la  M''(p'') | \bar{q}''_1 (k_1'')\Gamma q'_1(k_1') |M'(p') \ra \,,
\end{align}
 which will be further used to extract the form factors. For the detailed theoretical frameworks of SLF and CLF QMs, one  may refer to, for instance,  Refs.~\cite{Jaus:1989au,Jaus:1989av,Cheng:1996if} and Refs.~\cite{Jaus:1999zv,Cheng:2003sm}, respectively.

\subsection{The SLF quark model }
In the framework of SLF QM, a meson bound-state  consisting a quark $q_1$ and antiquark $\bar{q}_2$ with a total momentum $p$ can be written as
\begin{align}
|M(p)\ra =  \sum_{h_1,h_2} \int \frac{\d^3 \tilde{k}_1}{(2\pi)^32\sqrt{k_1^+}} \frac{\d^3 \tilde{k}_2}{(2\pi)^32\sqrt{k_2^+}} (2\pi)^3 \delta^3 ({\tilde{p}-\tilde{k}_1-\tilde{k}_2}) \Psi_{h_1,h_2}(\tilde{k}_1,\tilde{k}_2)|q_1({k}_1,h_1)\ra|\bar{q}_2({k}_2,h_2)\ra\,,
\label{eq:Fockexp}
\end{align}
where, $\tilde{p}=(p^+,\mathbf{p_\bot})$ and  $\tilde{k}_{1,2}=(k_{1,2}^+,\mathbf{k}_{1,2\bot})$ are the on-mass-shell LF momenta, $\Psi_{h_1,h_2}(\tilde{k}_1,\tilde{k}_2)$ is the momentum-space wavefunction~(WF), and the one particle states are defined as 
\begin{align}
&|q_1({k}_1,h_1)\ra=\sqrt{2k_1^+}\,b_{h1}^{\dagger}(k_1)|0\ra\,,\quad |\bar{q}_2({k}_2,h_2)\ra=\sqrt{2k_2^+}\,d_{h2}^{\dagger}(k_2)|0\ra,\\
&\{b^{\dagger}_{h} (k), b_{h'} (k') \}= \{d^{\dagger}_{h} (k), d_{h'} (k') \}=(2\pi)^3  \delta(k^+-k'^{+})\delta^2({\bf k}_{\bot}-{\bf k}'_{\bot}) \delta_{h h'}\,.
\end{align}
The momenta of $q_1$ and $\bar{q}_{2}$ can be written in terms of the internal  LF relative momentum variables $(x,{\mathbf{ k}_{\bot}})$ as 
\begin{align}\label{eq:momk1}
k_1^+=xp^+\,,\quad\, \mathbf{k}_{1\bot}=x\mathbf{p}_{\bot}+\mathbf{k}_{\bot} \,,\qquad  k_2^+=\bar{x}p^+ \,,\quad\, \mathbf{k}_{2\bot}=\bar{x}\mathbf{p}_{\bot}-\mathbf{k}_{\bot}\,,
\end{align}
where, $\bar{x}=1-x$, $\mathbf{k}_{\bot}=(k^x\,,k^y)$ and $\mathbf{p}_{\bot}=(p^x\,,p^y)$.

The momentum-space WF $ \Psi_{h_1,h_2}(x, { \mathbf{k}_{\bot}})$  in Eq.~\eqref{eq:Fockexp} satisfies the normalization condition
and can be expressed as
\begin{align}
\label{eq:LFWFP2}
\Psi_{h_1,h_2}(x,\mathbf{k}_{\bot})=S_{h_1,h_2}(x,\mathbf{k}_{\bot}) \psi(x,\mathbf{k}_{\bot}) \,,
\end{align}
where, $\psi(x,\mathbf{k}_{\bot})$ is the radial WF and responsible for describing the momentum distribution of the constituent quarks in the bound-state; $S_{h_1,h_2}(x,\mathbf{k}_{\bot})$  is the spin-orbital WF and responsible for constructing a state of definite spin $(S,S_z)$ out of the LF helicity $(h_1,h_2)$ eigenstates.  For the former, we shall use the Gaussian-type WF 
\begin{align}
\label{eq:RWFs}
\psi_s(x,\mathbf{k}_{\bot}) =4\frac{\pi^{\frac{3}{4}}}{\beta^{\frac{3}{2}}} \sqrt{ \frac{\partial k_z}{\partial x}}\exp\left[ -\frac{k_z^2+\mathbf{k}_\bot^2}{2\beta^2}\right]\,,
\end{align}
in this paper, where $k_z$ is the relative momentum in $z$-direction and has the form
\begin{align}
 k_z=(x-\frac{1}{2})M_0+\frac{m_2^2-m_1^2}{2 M_0}\,,
\end{align}
with the invariant mass
\begin{align}
M_0^2=\frac{m_1^2+\mathbf{k}_{\bot}^2}{x}+\frac{m_2^2+\mathbf{k}_{\bot}^2}{\bar{x}}\,.
\end{align}

The spin-orbital WF, $S_{h_1,h_2}(x,\mathbf{k}_{\bot}) $, can be obtained by the interaction-independent Melosh transformation.  It is convenient to use the covariant form, which can be further reduced by using the equation of motion on spinors and finally written as~\cite{Jaus:1989av,Cheng:2003sm}
\begin{align}\label{eq:defS2}
S_{h_1,h_2}=\frac{\bar{u}(k_1,h_1)\Gamma' v(k_2,h_2)}{\sqrt{2}\hat{M}_0}
\end{align}
where $\hat{M}_0^2 \equiv M_0^2-(m_1-m_2)^2$ and 
\begin{align}
&\Gamma'_P=\r_5\,,\\
& \Gamma'_V=-\not\!\hat{\epsilon}+\frac{\hat{\epsilon}\cdot (k_1-k_2)}{D_{ V,{\rm LF}}}\,,\quad D_{V,{\rm LF}}=M_0+m_1+m_2\,,
\label{eq:vSLF}
\end{align}
with
\begin{align}
\hat{\epsilon}^{\mu}_{\lbd=0}=&\frac{1}{M_0}\left(p^+,\frac{-M_0^2+\mathbf{p}_{\bot}^2}{p^+},\mathbf{p}_{\bot}\right)\,,\\
\hat{\epsilon}^{\mu}_{\lbd=\pm}=&\left(0,\frac{2}{p^+}\boldsymbol{\epsilon}_{\bot}\cdot \mathbf{p}_{\bot}, \boldsymbol{\epsilon}_{\bot}\right)\,,
\quad \boldsymbol{\epsilon}_{\bot}\equiv \mp \frac{(1,\pm i)}{\sqrt{2}}\,.
\end{align}



 In practice, for $M'(p')\to M''(p'')$ transition, we shall take the convenient Drell-Yan-West frame, $q^+=0$,  where $q\equiv p'-p''=k_1'-k_1''$ is the momentum  transfer. It implies that the form factors are known only for space-like momentum transfer, $q^2=-\mathbf{q}_\bot^2\leqslant 0$, and the ones in the time-like region need  an additional $q^2$ extrapolation. In addition,  we also take  a Lorentz frame where $\mathbf{p}_{\bot}'=0$ and $\mathbf{p}_{\bot}''=-\mathbf{q}_{\bot}$  amounts to $\mathbf{k}_{\bot}''=\mathbf{k}_{\bot}'-\bar{x}\mathbf{q}_{\bot}$. 
Finally, equipping Eq.~\eqref{eq:amp1} with the formulae given above and making some simplification, we obtain
\begin{eqnarray}
{\cal B}(q^2)=\sum_{h'_1,h''_1,h_2} \int  \frac{\d x \,\d^2{ \mathbf{k}_\bot'}}{(2\pi)^3\,2x}  {\psi''}^{*}(x,\mathbf{k}_{\bot}''){\psi'}(x,\mathbf{k}_{\bot}')
S''^{\dagger}_{h''_1,h_2}(x,\mathbf{k}_{\bot}'')\,C_{h''_1,h'_1}(x,\mathbf{k}_{\bot}',\mathbf{k}_{\bot}'')\,S'_{h'_1,h_2}(x,\mathbf{k}_{\bot}')\,,
\label{eq:B}
\end{eqnarray}
where $C_{h''_1,h'_1}(x,\mathbf{k}_{\bot}',\mathbf{k}_{\bot}'') \equiv  \bar{u}_{h''_1}(x,\mathbf{k}_{\bot}'')  \Gamma   u_{h'_1}(x,\mathbf{k}_{\bot}')$.

\subsection{The CLF quark model }
\begin{figure}[t]
\caption{The Feynman diagram for the matrix element $\cal B$.}
\begin{center}
\includegraphics[scale=0.3]{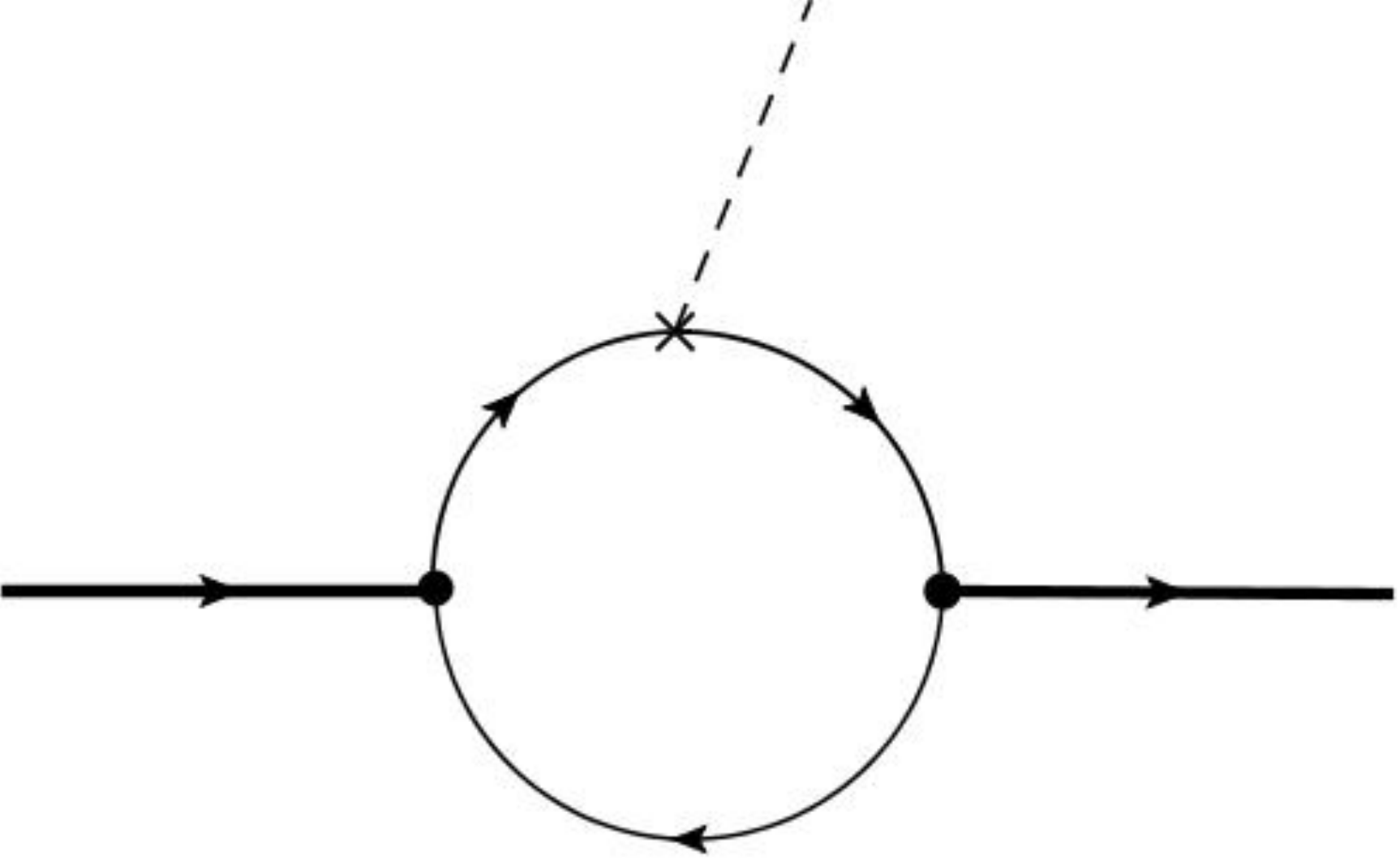}
\end{center}
\label{fig:fayn}
\end{figure}

In the CLF QM, the matrix element $\cal B$ is obtained by calculating the Feynman diagrams shown in Fig.~\ref{fig:fayn}. From this Feynman diagram and using the Feynman rules given in Refs.~\cite{Jaus:1999zv,Cheng:2003sm}, the matrix element ${\cal B}(q^2)$ can be written as a manifest covariant form,
\begin{eqnarray}\label{eq:Bclf1}
{\cal B}=N_c \int \frac{\d^4 k_1'}{(2\pi)^4} \frac{H_{M'}H_{M''}}{N_1'\,N_1''\,N_2}iS_{\cal B}\,,
\end{eqnarray}
where $\d^4 k_1'=\frac{1}{2} \d k_1'^- \d k_1'^+ \d^2 \mathbf{k}_{\bot}'$, the denominators $N_{1}^{(\prime,\prime\prime)}=k_{1}^{(\prime,\prime\prime)2}-m_1^{(\prime,\prime\prime)2}+i\e$ and $N_{2}=k_{2}^{2}-m_2^{2}+i\e$  come from the fermion propagators, and $H_{M', M''}$ are the bound-state vertex functions. The trace term  $S_{\cal B}$  associated with the fermion loop is written as 
\begin{eqnarray}
S_{\cal B}={\rm Tr}\left[\Gamma\, (\not\!k'_1+m'_1)\,(i\Gamma_{M'})\,(-\!\not\!k_2+m_2)\,(i\r^0{\Gamma}_{M''}^{\dag}\r^0) (\not\!k_1''+m_1'')\right]\,,
\end{eqnarray}
where the vertex operators $\Gamma_{M'}$ and ${\Gamma}_{M''}$ are relevant to the types of mesons and have the forms~\cite{Cheng:2003sm}
\begin{eqnarray}
&&i\Gamma_P=-i\r_5\,, \qquad i\Gamma_V=i\left[\gamma^\mu-\frac{ (k_1-k_2)^\mu}{D_{ V,{\rm con}}}\right]\,,
\end{eqnarray}
for $P$ and $V$ mesons, respectively. 

Integrating out the minus components of the loop momentum, one goes from the covariant calculation to the LF one. By closing the
contour in the upper complex $k_1'^-$ plane and assuming that $H_{M', M''}$ are analytic within the contour, the integration picks up a residue at $k_2^2=\hat{k}_2^2=m_2^2$ corresponding to put the spectator antiquark on the mass shell. Consequently, one has the following replacements~\cite{Jaus:1999zv,Cheng:2003sm}
\begin{eqnarray}
N_1 \to \hat{N}_1=x \left( M^2-M_0^2\right)
\end{eqnarray}
and
\begin{eqnarray}\label{eq:type1}
\chi_M = H_M/N\to h_M/\hat{N}\,,\qquad  D_{M,{\rm  con}} \to D_{M,{\rm  LF}}\,,\qquad \text{(type-I)}
\end{eqnarray}
where the LF forms of vertex function, $h_M$,  is given by
\begin{eqnarray}
h_P/\hat{N}&=&h_V/\hat{N}=\frac{1}{\sqrt{2N_c}}\sqrt{\frac{\bar{x}}{x}}\frac{\psi}{\hat{M}_0}\,.
\label{eq:vPV}
\end{eqnarray}
Eq.~\eqref{eq:type1} gives the correspondence between the manifest covariant  and LF approaches. As has been detailed in Ref.~\cite{Jaus:1999zv,Cheng:2003sm}, the correspondence between $\chi$ and $\psi$ in Eq.~\eqref{eq:type1} can be clearly derived by matching the CLF expressions to the SLF ones for some zero-mode independent  quantities, such as  $f_P$ and  $f_{+}^{P\to P}(q^2)$. However, the validity of the correspondence for the $D$ factor appearing in the vertex operator, $D_{M,{\rm  con}} \to D_{M,{\rm  LF}}$, has not yet been clarified explicitly~\cite{Choi:2013mda}. Instead of the traditional type-I correspondence, a much more generalized replacement,
\begin{eqnarray}\label{eq:type2}
\chi_M = H_M/N\to h_M/\hat{N}\,,\qquad  M\to  M_0\,,\qquad \text{(type-II)}
\end{eqnarray}
is suggested for  the purpose of  self-consistence of $f_{A,V}$~\cite{Choi:2013mda,Chang:2018zjq}. Our following theoretical results are given within traditional type-I scheme unless otherwise specified. The ones within type-II scheme can be easily obtained by making an additional replacement $M\to  M_0$.
Finally, after integrating out $k_1'^-$, the matrix element, Eq.~\eqref{eq:Bclf1}, can be reduced as the LF form
\begin{eqnarray}
\label{eq:Bclf2}
\hat{{\cal B}}=N_c \int \frac{\d x \d^2 \mathbf{k}_{\bot}'}{2(2\pi)^3}\frac{h_{M'}h_{M''}}{\bar{x} \hat{N}_1'\,\hat{N}_1''\,}\hat{S}_{\cal B}\,.
\end{eqnarray}

It has been noted in Refs.~\cite{Jaus:1999zv,Cheng:2003sm} that ${\cal B}$ receives additional spurious contributions proportional to the light-like vector $\omega^\mu=(0,2,\mathbf{0}_\bot)$, and these undesired spurious contributions are expected to be cancelled out by the zero-mode contributions. As demonstrated in  Ref.~\cite{Jaus:1999zv}, the inclusion of the zero mode contribution in practice amounts to some proper replacements in the $\hat{S}_{\cal B}$ under integration. For the quantities studied in this paper, we need~\cite{Jaus:1999zv,Cheng:2003sm}
\begin{align}
\hat{k}_1'^{\mu} &\to P^\u A_1^{(1)}+q^\u A_2^{(1)} \,,\nonumber\\
\hat{k}_1'^{\mu}\hat{k}_1'^{\nu} &\to g^{\u\v}A_1^{(2)}+P^\u P^\v A_2^{(2)}+(P^\u q^\v+q^\u P^\v)A_3^{(2)}+q^\u q^\v A_4^{(2)}\nonumber\\
&~~~~+\frac{P^\u\omega^\v+\omega^\u P^\v}{\omega\cdot P}B_1^{(2)}\,,\nonumber\\
k_1'^{\mu}\hat{N}_2&\to q^\u\left(A_2^{(1)}Z_2+\frac{q\cdot P}{q^2}A_1^{(2)} \right) \,,\nonumber\\
\hat{k}_1'^{\mu}\hat{k}_1'^{\nu}\hat{N}_2&\to g^{\u\v}A_1^{(2)}Z_2+q^\u q^\v\left( A_4^{(2)}Z_2+2\frac{q\cdot P}{q^2}A_2^{(1)}A_1^{(2)}\right)+\frac{P^\u\omega^\v+\omega^\u P^\v}{\omega\cdot P}B_3^{(3)}\,,\nonumber\\
Z_2&=\hat{N}_1'+m_1'^2-m_2^2+(\bar{x}-x)M'^2+(q^2+q\cdot P)\frac{k_{1\bot}'\cdot q_{\bot}}{q^2}\,,
 \label{eq:repFF}
\end{align}
where $P=p'+p''$, and the $A$ and $ B$ functions are given by
\begin{align}
A_1^{(1)}&=  \frac{x}{2}\,,\qquad
A_2^{(1)}=\frac{x}{2} -\frac{k_{1\bot}' \cdot q_{\bot}}{q^2}\,,\qquad\nonumber\\
A_1^{(2)}&=-k_{1\bot}'^2 -\frac{(k_{1\bot}' \cdot q_{\bot})^2}{q^2}\,,
\nonumber\\
A_2^{(2)}&=(A_1^{(1)})^2\,,\qquad A_3^{(2)}=A_1^{(1)}A_2^{(1)}\,,\qquad
A_4^{(2)}=(A_2^{(1)})^2\,,\nonumber\\
B_1^{(2)}&=\frac{x}{2}Z_2-A_1^{(2)}\,,\qquad B_3^{(3)}=B_1^{(2)}Z_2+
\left(P^2-\frac{(q\cdot P)^2}{q^2}\right)A_1^{(1)}A_1^{(2)}. \label{eq:ABfuns}
\end{align}
It should be noted that most of the $\w$-dependent terms associated with the  $C$ functions have been eliminated by the inclusion of the zero-mode contributions~\cite{Jaus:1999zv}, and thus are not shown in above formulae. However, there are still some residual $\w$-dependences that are  associated with the  $B$ functions, which can be clearly seen from~Eq.~\eqref{eq:repFF}. As stated in Ref.~\cite{Jaus:1999zv}, the $B$ functions  play a special role since, on the one hand, it is combined with $\w^\u$, on the other hand, there is no zero-mode contribution associated with $B$ due to $x\hat{N}_2=0$.  Therefore, a different mechanism is required to neutralize the residual $\w$-dependence.

Using the formulae given above, one can obtain the full result of $\cal B$, and further extract the form factors. For a given quantity,~${\cal Q}$, its full result can be expressed as the sum of the valence and zero-mode contributions,
\begin{align}
{\cal Q}^{\rm full}={\cal Q}^{\rm val.}+{\cal Q}^{\rm z.m.}\,.
\end{align}
In order to evaluate the effect of zero-mode, we also need to calculate ${\cal Q}^{\rm val.}$ and/or ${\cal Q}^{\rm z.m.}$.  In this paper, we employ the strategy introduced in Ref.~\cite{Chang:2018zjq} to calculate ${\cal Q}^{\rm val.}$.

\section{Results and discussions}
The matrix element for the $P\to V$ transition can be represented in terms of the form factors as
\begin{align}
\label{eq:g}
\la V(p^{\prime\prime},\lambda) | \bar q_1^{\prime\prime}\r_\u q_1^\prime|P(p^\prime)\ra&=i\varepsilon_{\u\v\a\b}{\e}^{*\v}P^{\a}q^{\b}\,g(q^2)  \,,\\
\label{eq:fa}
\la V(p^{\prime\prime},\lambda) | \bar q_1^{\prime\prime}\r_\u\r^5 q_1^\prime|P(p^\prime)\ra&=-f(q^2)\, {\e}^*_{\u}-{\e}^*\cdot P \left[a_+(q^2) P_\u +a_-(q^2) q_\u \right]\,.
\end{align}
These form factors are related to the commonly used Bauer-Stech-Wirbel~(BSW) form factors via
 \begin{align}
 &V(q^2)=-(M'+M'') g(q^2)\,,\quad A_1(q^2)=-\frac{ f(q^2)}{M'+M''}\,,\quad A_2(q^2)=(M'+M'') a_+(q^2)\,,\nonumber\\
 &A_0(q^2)=-\frac{1}{2M''}\left[q^2 a_-(q^2) +f(q^2)+(M'^2-M''^2)a_+(q^2)\right]\,.
 \end{align}

\subsection{Theoretical results}
Using the formulae given in the last section, we obtain the SLF results for the form factors written as
\begin{align}
[{\cal F}(q^2)]_{\rm SLF}=\int  \frac{\d x\, \d^2{\bf k}_{\bot}'}{(2\pi)^3\,2x}  \frac{ {\psi''}^{*}(x,{{\bf k}}_{\bot}''){\psi'}(x,{ { \bf k}}_{\bot}') }{2\hat{M}'_0\hat{M}''_0}\,\widetilde{\cal F}^{\rm SLF}(x,{\bf k}_{\bot}',q^2),
\end{align} 
where, ${\cal F}=g$, $f$ and $a_\pm$, and the integrands are
\begin{align}
\label{eq:gSLF}
\widetilde{g}^{\rm SLF}(x,{\bf k}_{\bot}',q^2)=&-2\left\{\bar{x}m_1'+xm_2+(m_1'-m_1'') \frac{{\bf k}_{\bot}'\cdot {\bf q}_{\bot}}{q^2}+\frac{2}{D_{\rm V, LF}''}\left[{ \bf k}_{\bot}'^2+\frac{({\bf k}_{\bot}'\cdot {\bf q}_{\bot})^2}{q^2}\right]\right\}\,,\\
 \label{eq:apSLF}
  \widetilde{a}_{+}^{\rm SLF}(x,{\bf k}_{\bot}',q^2)=&2 \bigg\{  (m_1''-2xm_1'+m_1'+2xm_2)\, \frac{ {\bf k}_{\bot}'\cdot {\bf q}_{\bot }}{{\bf q}_{\bot }^2} + ( x-\bar{x})(\bar{x}m_1'+x m_2)   \nonumber\\
& +\frac{2}{D_{\rm V, LF}''}  \left( \frac{{\bf k}_{\bot}'\cdot {\bf q}_{\bot}}{\bar{x} {\bf q}_{\bot}^2} -1\right)  \left[ {\bf k}_{\bot}'\cdot {\bf k}_{\bot}''+  (x m_2 - \bar{x}m_1'')(x m_2+ \bar{x}m_1') \right] \bigg\} \,,\\
 \label{eq:fSLF}
\widetilde{f}^{\rm SLF}(x,{\bf k}_{\bot}',q^2)=&- \frac{4M''}{\bar{x}M_0''} \bigg\{ \Big[  {\bf k}_{\bot}'^2(\bar{x}m_1'+m_1''-\bar{x}m_2) - \bar{x} {\bf k}_{\bot}'\cdot {\bf q}_{\bot}(m_1''+2\bar{x}m_1'+xm_2-\bar{x}m_2)\nonumber\\
 &+ (\bar{x}m_1' + x m_2) (m_1''m_2+x\bar{x} M_0''^2+\bar{x}^2{\bf q}_{\bot}^2) \Big]\nonumber\\
 & + \frac{ {\bf k}_{\bot}''^2+m_2^2-\bar{x}^2M_0''^2 }{\bar{x} D_{\rm V, LF}''} \big[  {\bf k}_{\bot}'\cdot {\bf k}_{\bot}''+  (x m_2 - \bar{x}m_1'')(x m_2+ \bar{x}m_1')  \big] \bigg\}\nonumber\\
&  - \left(M'^2 -M''^2+{\bf q}_\bot^2 \right)\,  \widetilde{a}_{+}^{\rm SLF}(x,{\bf k}_{\bot}',q^2)\,,\\
 \label{eq:amSLF}
\widetilde{a}_{-}^{\rm SLF}(x,{\bf k}_{\bot}',q^2)=& \frac{4}{{\bf q}_{\bot}^2} \bigg\{
m_1''M_0'^2+m_1'M_0''^2 -(m_1'+m_1'') (m_1'-m_2) (m_1''-m_2)\nonumber \\
  &-\bar{x}(m_1'-m_2) {\bf q}_{\bot}^2+[m_1'-m_1''+2\bar{x}(m_1'-m_2)] {\bf k}_{\bot}'\cdot {\bf q}_{\bot}-2(m_1'-m_2){\bf k}_{\bot}'^2 \nonumber \\
&+\frac{1}{D_{\rm V, LF}''}\Big\{- {\bf k}_{\bot}''\cdot {\bf q}_{\bot}\left[ M_0'^2-(m_1'-m_2)^2 \right] \nonumber \\
&+ {\bf k}_{\bot}'\cdot {\bf k}_{\bot}'' \left[ M_0''^2  +M_0'^2+2 (m_1'' + m_2) (m_1' - m_2)+{\bf q}_{\bot}^2\right] \Big\}
\bigg\}\nonumber\\
&-\frac{2}{q^2}\widetilde{f}^{\rm SLF}(x,{\bf k}_{\bot}',q^2)+   \widetilde{a}_{+}^{\rm SLF}(x,{\bf k}_{\bot}',q^2)\,.
\end{align}
The SLF results for  $g(q^2)$ and $a_+(q^2)$ have been given in the previous works, for instance, Refs.~\cite{Jaus:1989au,Jaus:1989av}, while the ones for $f(q^2)$ and $a_-(q^2)$ are first obtained in this paper.  For $[g(q^2)]_{\rm SLF}$ and $[a_+(q^2)]_{\rm SLF}$, the results in Ref.~\cite{Jaus:1989av} are obtained by taking ${\bf p}_{\bot}''=0$~({\it i.e.} the final state moves along the $z$ axis), while the  results given above and the ones in Ref.~\cite{Jaus:1989au} are obtained in the ${\bf p}_{\bot}'=0$ reference frame~({\it i.e.} the initial state moves along the $z$ axis). 
We find that these two sets of results  are a little bit different in form, but such difference does not affect the final results after the internal variables are integrated out.  In addition, it is claimed in the abstract of Ref.~\cite{Jaus:1999zv} that $a_-(q^2)$ is impossible to be calculated in the SLF QM,  which will be discussed in the next subsection.

Using the framework  of the CLF QM given in the last section, we obtain the full CLF results for the form factors, 
\begin{align}\label{eq:Ffull}
[{\cal F}(q^2)]_{\rm full}=N_c \int \frac{\d x \,  \d^2{\bf k}_{\bot}'}{2(2\pi)^3}\frac{\chi_P'\chi_V''}{\bar{x}}\,\widetilde{\cal F}^{\rm full}(x,{\bf k}_{\bot}',q^2)\,,
\end{align} 
where, 
\begin{align} 
\label{eq:gfull}
\widetilde{g}^{\rm full}(x,{\bf k}_{\bot}',q^2)=&-2\left\{\bar{x} m_1' +xm_2+(m_1' - m_1'' )\frac{{\bf k}_{\bot}'\cdot {\bf q}_\bot}{q^2}+\frac{2}{D_{\rm V, con}''}\left[{\bf k}_{\bot}'^2 +\frac{({\bf k}_{\bot}' \cdot {\bf q}_{\bot})^2}{q^2} \right] \right\}\,,\\
 \label{eq:apfull}
 \widetilde{a}_{+}^{\rm full}(x,{\bf k}_{\bot}',q^2)
=&2 \bigg\{  (m_1''-2xm_1'+m_1'+2xm_2)\, \frac{ {\bf k}_{\bot}'\cdot {\bf q}_{\bot }}{{\bf q}_{\bot }^2} + ( x-\bar{x})(\bar{x}m_1'+x m_2)   \nonumber\\
& +\frac{2}{D_{\rm V, con}''}   \frac{{\bf k}_{\bot}''\cdot {\bf q}_{\bot}}{\bar{x} {\bf q}_{\bot}^2}  \left[ {\bf k}_{\bot}'\cdot{\bf k}_{\bot}''+  (x m_2 - \bar{x}m_1'')(x m_2+ \bar{x}m_1') \right] \bigg\} \,,\\
 \label{eq:ffull}
\widetilde{f}^{\rm full}(x,{\bf k}_{\bot}',q^2)=&-2\Bigg\{
-(m_1'+m_1'')^2 (m_1'-m_2) +(xm_2-\bar{x}m_1') M'^2+(xm_2+\bar{x}m_1') M''^2\nonumber\\ 
 & -x(m_2-m_1' )(M_0'^2+M_0''^2)+2xm_1''M_0'^2  -4 \left(m_1'-m_2\right) \left({\bf k}_{\bot}'^2 +\frac{({\bf k}_{\bot}' \cdot {\bf q}_{\bot})^2}{q^2}\right)\nonumber\\ 
 &- m_2 q^2 -(m_1'+m_1'')(q^2+q\cdot P)\frac{{\bf k}_{\bot}'\cdot {\bf q}_{\bot}}{q^2} {+4 (m_1'-m_2)  B_1^{(2)}}\nonumber\\ 
 &+\frac{2}{D_{\rm V, con}''} \bigg[ \left({\bf k}_{\bot}'^2 +\frac{({\bf k}_{\bot}' \cdot {\bf q}_{\bot})^2}{q^2}\right) \bigg((x-\bar{x})M'^2+M''^2-2(m_1'-m_1'') (m_1'-m_2) \nonumber\\ 
&+2xM_0'^2-q^2 -2 (q^2+q\cdot P)\frac{{\bf k}_{\bot}'\cdot {\bf q}_{\bot}}{q^2}\bigg) \nonumber\\ 
 &- \left(M'^2+M''^2-q^2 + 2(m_1'-m_2)(m_1''+m_2)\right) B_1^{(2)} +2  B_3^{(3)}\bigg]\Bigg\}\,,\\
  \label{eq:amfull}
\widetilde{a}_{-}^{\rm full}(x,{\bf k}_{\bot}',q^2)=&-2\Bigg\{   (3-2x)(\bar{x}m_1'+xm_2) -  \left[(6x-7)m_1'+(4-6x)m_2+m_1''\right] \frac{{\bf k}_{\bot}' \cdot {\bf q}_{\bot}}{q^2}\nonumber\\ 
  & +4(m_1'-m_2)\left[2\left(\frac{{\bf k}_{\bot}' \cdot {\bf q}_{\bot}}{q^2}\right)^2+\frac{{\bf k}_{\bot}'^2}{q^2}\right] {- 4 \frac{(m_1'-m_2)  }{q^2}B_1^{(2)}}\nonumber\\ 
 &+\frac{1}{D_{\rm V, con}''} \bigg[-2\left(M'^2+M''^2-q^2 + 2(m_1'-m_2)(m_1''+m_2)\right) (A_3^{(2)}+A_4^{(2)}-A_2^{(1)})\nonumber\\ 
 &+ \left( 2M'^2-q^2-\hat{N}_1'+\hat{N}_1''- 2(m_1'-m_2)^2+(m_1' +m_1'')^2 \right) \left( A_1^{(1)}+ A_2^{(1)} -1\right)\nonumber\\ 
&+2Z_2 \left( 2A_4^{(2)}-3A_2^{(1)} +1\right) + 2 \frac{q\cdot P}{q^2} \left( 4 A_2^{(1)}A_1^{(2)} - 3A_1^{(2)} \right) \nonumber\\ 
&{+\frac{2}{q^2}  \left(\left(M'^2+M''^2-q^2 + 2(m_1'-m_2)(m_1''+m_2)\right) B_1^{(2)} -2  B_3^{(3)}\right)}\bigg] \Bigg\}\,,
\end{align}
for the case of  $\lbd=0$~({\it i.e.}, $[\widetilde{{\cal F}}^{\rm full}]^{\lbd=0}$); and the results for the case of  $\lbd=\pm$~({\it i.e.}, $[\widetilde{{\cal F}}^{\rm full}]^{\lbd=\pm}$) can be obtained from these formulas by deleting  the  terms  associated with $B$ functions. The CLF results for ${\cal F}_{P\to V}$  are also given in Refs.~\cite{Jaus:1999zv,Cheng:2003sm}, but the contributions associated with $B$ functions are not considered.  In Ref.~\cite{Cheng:2003sm}, the authors claim that  the contributions of $B$ functions to form factors vanish when taking $\lbd=\pm$, which is also checked carefully in this work and found to be legitimate. However, we find that the contributions of $B$ functions always exist for the $\lbd=0$ state,
which implies that  $f(q^2)$ and $a_-(q^2)$ possibly suffer from the problem of self-consistency like the case of $f_V$ found in Refs.~\cite{Cheng:2003sm,Choi:2013mda,Chang:2018zjq}. In fact, we will show later that whether $[{\cal F}_{P\to V}]^{\lbd=\pm}$  receive the contributions of $B$ functions is determined by the strategy employed to deal with the spurious $\w$ dependent contributions.

Meanwhile, the corresponding valence contributions in the CLF approach can also been obtained via. Eq.~\eqref{eq:Ffull} with the integrands 
\begin{align}
 \label{eq:gval}
\widetilde{g}^{\rm val.}(x,{\bf k}_{\bot}',q^2)=&\widetilde{g}^{\rm full}(x,{\bf k}_{\bot}',q^2)\,,\\
 \label{eq:apval}
  \widetilde{a}_{+}^{\rm val.}(x,{\bf k}_{\bot}',q^2)=& \widetilde{a}_{+}^{\rm full}(x,{\bf k}_{\bot}',q^2)\,,\\
  \label{eq:fval}
\widetilde{f}^{\rm val.}(x,{\bf k}_{\bot}',q^2)=&-\frac{4}{\bar{x}} \bigg\{  {\bf k}_\bot'^2(\bar{x}m_1'+m_1''-\bar{x}m_2)
-\bar{x}{\bf k}_\bot'\cdot {\bf q}_\bot ( m_1''+2\bar{x}m_1'+xm_2-\bar{x}m_2)\nonumber\\  
&+(\bar{x}m_1'+xm_2)\left[m_1''m_2+x\bar{x}M''^2+\bar{x}^2{\bf q}_{\bot}^2\right]  \nonumber\\ 
&+ \frac{{\bf k}_\bot''^2+m_2^2-\bar{x}^2M''^2}{\bar{x}D_{\rm V, con}''}\left[{\bf k}_\bot'\cdot {\bf k}_\bot''+(xm_2+\bar{x}m_1')(xm_2-\bar{x}m_1'')\right] \bigg\}   \nonumber\\   
 & -\left(M'^2-M''^2+{\bf q}_\bot^2\right) \widetilde{a}_{+}^{\rm val.}(x,{\bf k}_{\bot}',q^2)\,,\\
  \label{eq:amval}
\widetilde{a}_{-}^{\rm full}(x,{\bf k}_{\bot}',q^2)=&\frac{4}{{\bf q}_{\bot}^2}\bigg\{ x(m_1''-m_2)M_0'^2+x(m_1'-m_2)M_0''^2+(\bar{x}m_1''+xm_2)M'^2+(\bar{x}m_1'+xm_2)M''^2 \nonumber \\
  &-(m_1'+m_1'') (m_1'-m_2) (m_1''-m_2)-\bar{x}(m_1'-m_2) {\bf q}_{\bot}^2\nonumber\\   
  & +[m_1'-m_1''+2\bar{x}(m_1'-m_2)]{\bf k}_{\bot}'\cdot {\bf q}_{\bot}-2(m_1'-m_2){\bf k}_{\bot}'^2 \nonumber\\ 
 &  +\frac{1}{D_{\rm V, con}''}\Big[-{\bf k}_\bot''\cdot {\bf q}_\bot\left( xM_0'^2+\bar{x}M'^2-(m_1'-m_2)^2 \right) \nonumber \\
&+{\bf k}_\bot'\cdot {\bf k}_\bot'' \left(M''^2  +M'^2+2 (m_1'' + m_2) (m_1' - m_2)+{\bf q}_{\bot}^2\right) \Big] \bigg\} \nonumber\\
&+\frac{2}{{\bf q}_{\bot}^2}\widetilde{f}^{\rm val.}(x,{\bf k}_{\bot}',q^2)+\widetilde{g}^{\rm val.}(x,{\bf k}_{\bot}',q^2)\,.
\end{align}

\subsection{Numerical results and discussions}

\begin{table}[t]
\begin{center}
\caption{\label{tab:1} \small Numerical results of the zero-mode independent form factors  $g({\textbf{q}}_\perp^2)$ and $a_+({\textbf{q}}_\perp^2)$ at ${\bf q}_{\bot}^2=(0,2,4) \,{\rm GeV^2}$ for $B\to D^*$ transition and at ${\bf q}_{\bot}^2=(0,1,2) \,{\rm GeV^2}$ for $D\to K^*$ transition .}
\vspace{0.2cm}
\let\oldarraystretch=\arraystretch
\renewcommand*{\arraystretch}{1.1}
\setlength{\tabcolsep}{8.8pt}
\begin{tabular}{lcccccccccccc}
\hline\hline
     $B\to D^*$
                 &${\textbf{q}}_\perp^2=0$
                 &${\textbf{q}}_\perp^2=2$
                 &${\textbf{q}}_\perp^2=4$
                \\\hline
$g({\textbf{q}}_\perp^2)$
                 &$-0.11$&$-0.10$&$-0.09$\\\hline
$a_+({\textbf{q}}_\perp^2)$
                 &$0.08$&$0.08$&$0.07$
\\\hline
    $D\to K^*$
                 &${\textbf{q}}_\perp^2=0$
                 &${\textbf{q}}_\perp^2=1$
                 &${\textbf{q}}_\perp^2=2$
                \\\hline
$g({\textbf{q}}_\perp^2)$
                 &$-0.35$&$-0.27$&$-0.22$\\\hline
$a_+({\textbf{q}}_\perp^2)$
                 &$0.20$&$0.16$&$0.14$
\\\hline\hline
\end{tabular}
\end{center}
\end{table}
\begin{table}[t]
\begin{center}
\caption{\label{tab:2} \small Numerical results of the zero-mode dependent form factors  $f({\bf{q}}_\perp^2)$ and $a_-({\bf{q}}_\perp^2)$ at ${\bf q}_{\bot}^2=(0,2,4) \,{\rm GeV^2}$ for $B\to D^*$ transition.}
\vspace{0.2cm}
\let\oldarraystretch=\arraystretch
\renewcommand*{\arraystretch}{1.1}
\setlength{\tabcolsep}{8.8pt}
\begin{tabular}{l|cccccccccccc}
\hline\hline
$B\to D^*$    &  &$[f({\bf{q}}_\perp^2)]_{\text{SLF}}$
                 &$[f({\bf{q}}_\perp^2)]^{\lambda=0}_{\text{full}}$
                 &$[f({\bf{q}}_\perp^2)]^{\lambda=\pm1}_{\text{full}}$
                 &$[f({\bf{q}}_\perp^2)]_{\text{val.}}$\\\hline
\multirow{2}{*}{${\bf{q}}_\perp^2=0$}
&type-I     &$-4.82$&$-4.84$&$-5.01$&$-5.01$
              \\\cline{2-6}
&type-II    &$-5.32$&$-5.32$&$-5.32$&$-5.32$
\\\hline
\multirow{2}{*}{${\bf{q}}_\perp^2=2$}
&type-I     &$-4.64$&$-4.68$&$-4.78$&$-4.81$
              \\\cline{2-6}
&type-II    &$-5.11$&$-5.11$&$-5.11$&$-5.11$
\\\hline
\multirow{2}{*}{${\bf{q}}_\perp^2=4$}
&type-I     &$-4.48$&$-4.54$&$-4.58$&$-4.64$
              \\\cline{2-6}
&type-II    &$-4.92$&$-4.92$&$-4.92$&$-4.92$
\\\hline\hline
$B\to D^*$   &   &$[a_-({\bf{q}}_\perp^2)]_{\text{SLF}}$
                 &$[a_-({\bf{q}}_\perp^2)]^{\lambda=0}_{\text{full}}$
                 &$[a_-({\bf{q}}_\perp^2)]^{\lambda=\pm1}_{\text{full}}$
                 &$[a_-({\bf{q}}_\perp^2)]_{\text{val.}}$\\\hline
\multirow{2}{*}{${\bf{q}}_\perp^2=0$}
&type-I     &---
            &---
            &$-0.11$
            &---
              \\\cline{2-6}
&type-II    &$-0.10$&$-0.10$&$-0.10$&$-0.10$
\\\hline
\multirow{2}{*}{${\bf{q}}_\perp^2=2$}
&type-I     &$0.38$&$-0.87$&$-0.10$&$-0.13$
              \\\cline{2-6}
&type-II    &$-0.09$&$-0.09$&$-0.09$&$-0.09$
\\\hline
\multirow{2}{*}{${\bf{q}}_\perp^2=4$}
&type-I     &$0.14$&$-0.50$&$-0.09$&$-0.10$
              \\\cline{2-6}
&type-II    &$-0.08$&$-0.08$&$-0.08$&$-0.08$
\\\hline\hline
\end{tabular}
\end{center}
\end{table}
\begin{table}[t]
\begin{center}
\caption{\label{tab:3} \small Numerical results of the zero-mode dependent form factors  $f({\bf{q}}_\perp^2)$ and $a_-({\bf{q}}_\perp^2)$ at ${\bf q}_{\bot}^2=(0,1,2) \,{\rm GeV^2}$ for $D\to K^*$ transition . }
\vspace{0.2cm}
\let\oldarraystretch=\arraystretch
\renewcommand*{\arraystretch}{1.1}
\setlength{\tabcolsep}{8.8pt}
\begin{tabular}{l|cccccccccccc}
\hline\hline
$D\to K^*$    &  &$[f({\bf{q}}_\perp^2)]_{\text{SLF}}$
                 &$[f({\bf{q}}_\perp^2)]^{\lambda=0}_{\text{full}}$
                 &$[f({\bf{q}}_\perp^2)]^{\lambda=\pm1}_{\text{full}}$
                 &$[f({\bf{q}}_\perp^2)]_{\text{val.}}$\\\hline
\multirow{2}{*}{${\bf{q}}_\perp^2=0$}
&type-I     &$-1.93$&$-1.76$&$-2.17$&$-2.19$
              \\\cline{2-6}
&type-II    &$-2.66$&$-2.66$&$-2.66$&$-2.66$
\\\hline
\multirow{2}{*}{${\bf{q}}_\perp^2=1$}
&type-I     &$-1.75$&$-1.59$&$-1.89$&$-1.97$
              \\\cline{2-6}
&type-II    &$-2.37$&$-2.37$&$-2.37$&$-2.37$
\\\hline
\multirow{2}{*}{${\bf{q}}_\perp^2=2$}
&type-I     &$-1.61$&$-1.50$&$-1.69$&$-1.79$
              \\\cline{2-6}
&type-II    &$-2.14$&$-2.14$&$-2.14$&$-2.14$
\\\hline\hline
$D\to K^*$   &   &$[a_-({\bf{q}}_\perp^2)]_{\text{SLF}}$
                 &$[a_-({\bf{q}}_\perp^2)]^{\lambda=0}_{\text{full}}$
                 &$[a_-({\bf{q}}_\perp^2)]^{\lambda=\pm1}_{\text{full}}$
                 &$[a_-({\bf{q}}_\perp^2)]_{\text{val.}}$\\\hline
\multirow{2}{*}{${\bf{q}}_\perp^2=0$}
&type-I     
               &---
            &---
            &$-0.34$
            &---
              \\\cline{2-6}
&type-II    &$-0.35$&$-0.35$&$-0.35$&$-0.35$
\\\hline
\multirow{2}{*}{${\bf{q}}_\perp^2=1$}
&type-I     &$0.97$&$-0.94$&$-0.27$&$-0.49$
              \\\cline{2-6}
&type-II    &$-0.27$&$-0.27$&$-0.27$&$-0.27$
\\\hline
\multirow{2}{*}{${\bf{q}}_\perp^2=2$}
&type-I     &$0.32$&$-0.62$&$-0.22$&$-0.30$
              \\\cline{2-6}
&type-II    &$-0.21$&$-0.21$&$-0.21$&$-0.21$
\\\hline\hline
\end{tabular}
\end{center}
\end{table}

\begin{figure}[t]
\caption{The dependences of $\Delta^{f,a_-}_{\rm full}(x)$ and $\d[{f,a_-}]_{\rm z.m.}/\d x$ on $x$ for $B\to D^*$  transition at ${\bf q}_{\bot}^2=(0,2,4)\,{\rm GeV^2}$. See text for the detailed explanations and discussions.}
\begin{center}
\subfigure[]{\includegraphics[scale=0.1]{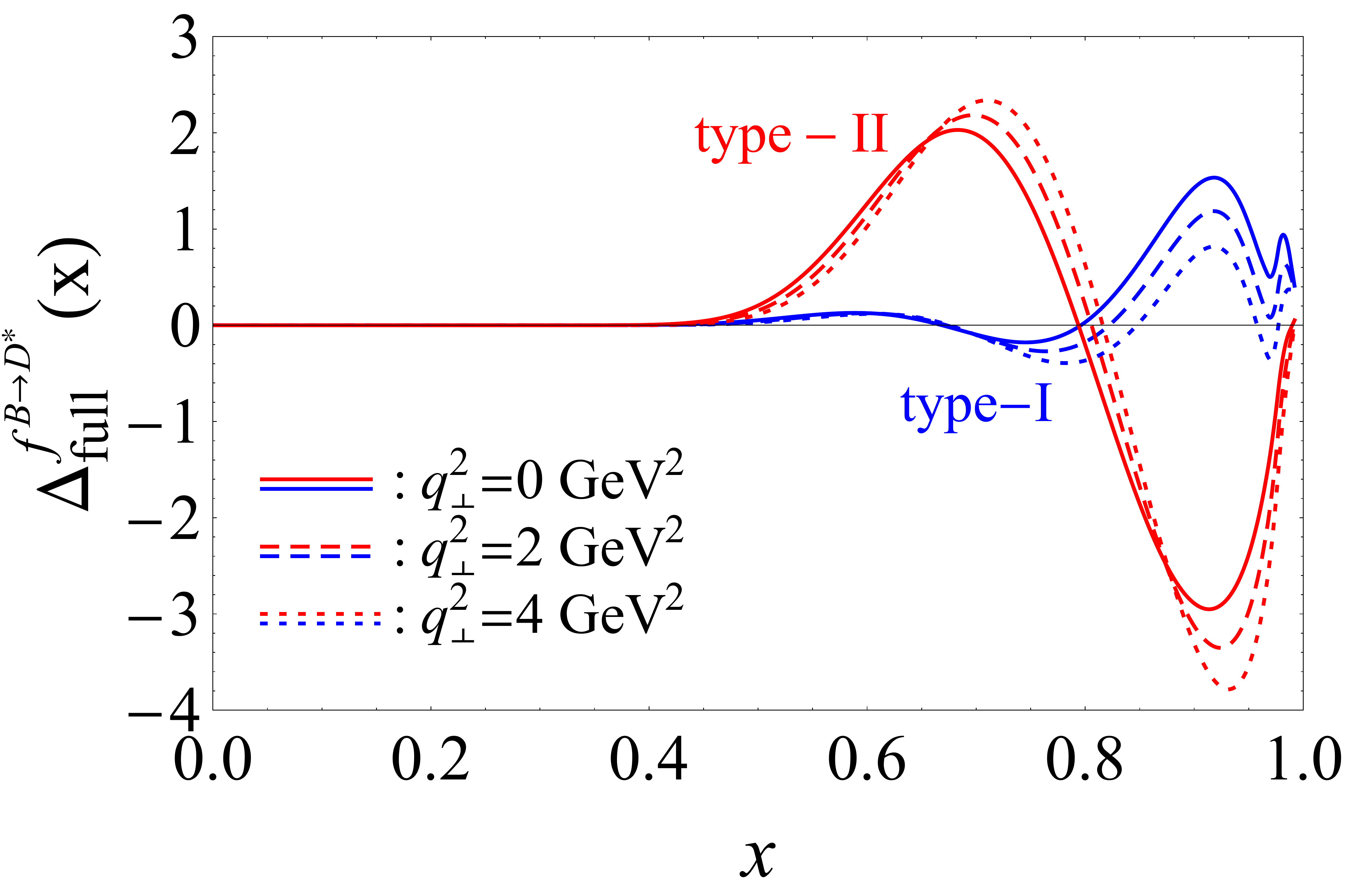}}\qquad\qquad
\subfigure[]{\includegraphics[scale=0.1]{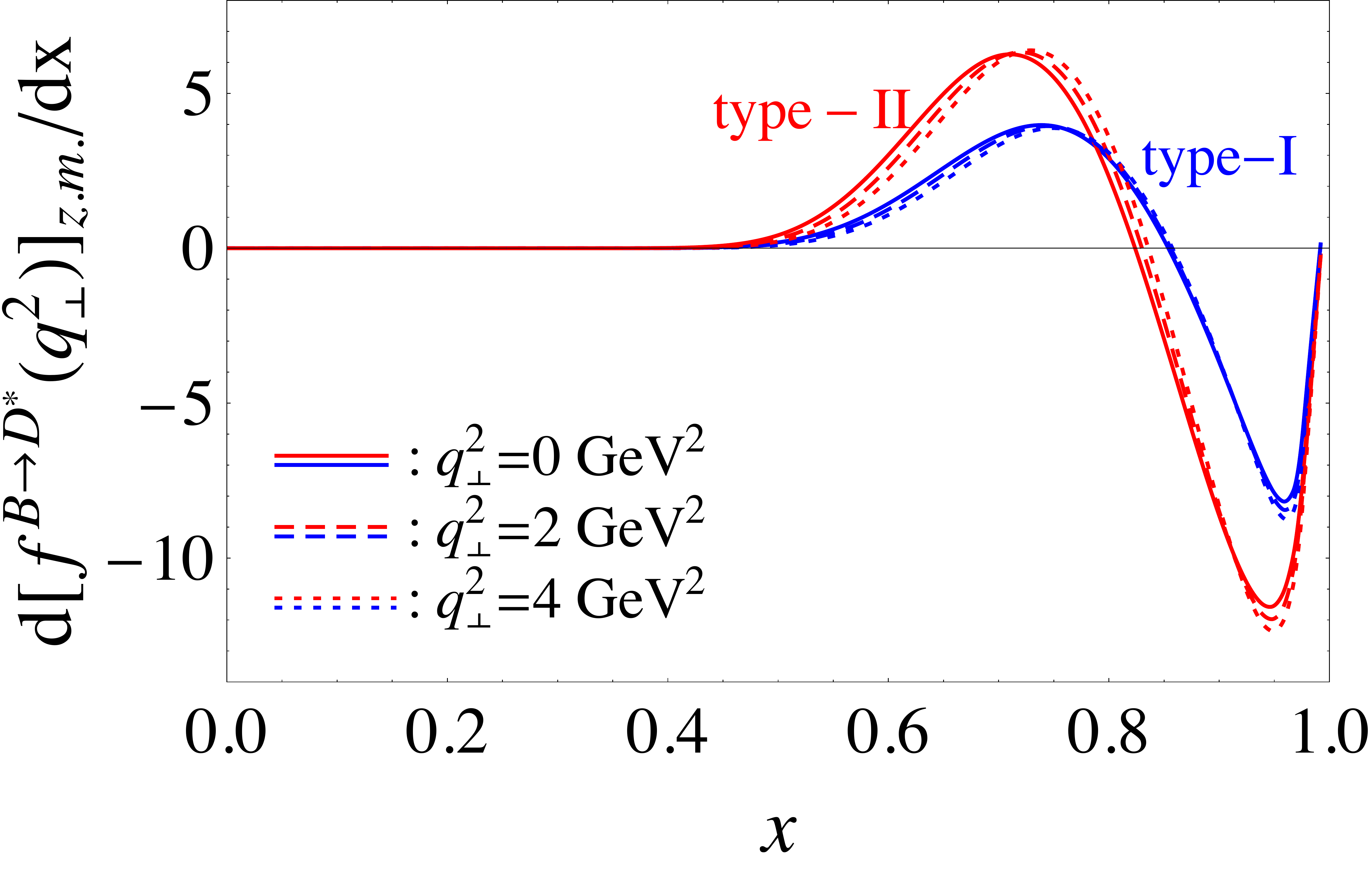}}\\
\subfigure[]{\includegraphics[scale=0.105]{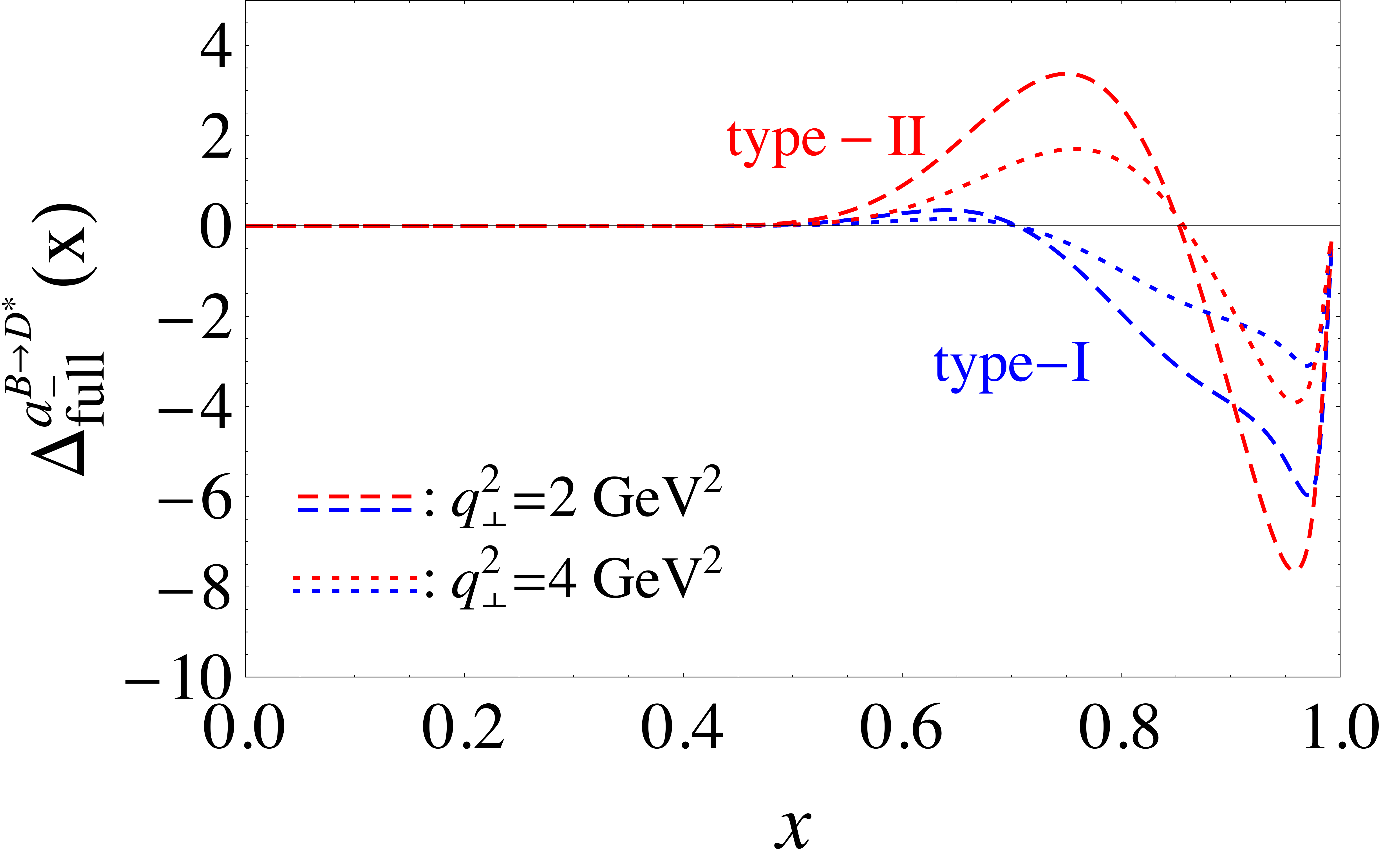}}\qquad\qquad
\subfigure[]{\includegraphics[scale=0.105]{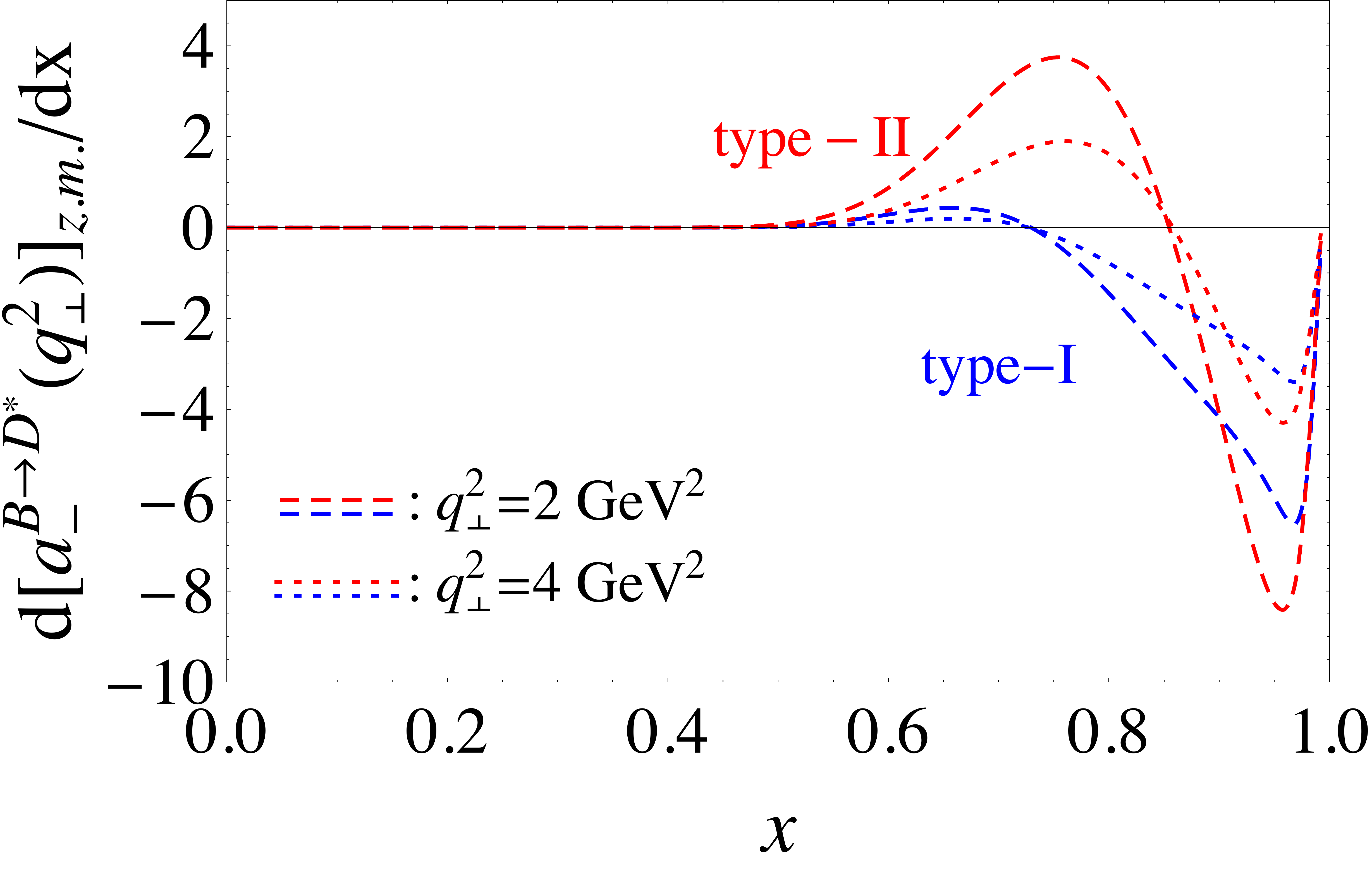}}
\end{center}
\label{fig:BD}
\end{figure}

\begin{figure}[t]
\caption{The dependences of $\Delta^{f,a_-}_{\rm full}(x)$ and $\d[{f,a_-}]_{\rm z.m.}/\d x$ on $x$ for $D\to K^*$  transition at ${\bf q}_{\bot}^2=(0,1,2)\,{\rm GeV^2}$. See text for the detailed explanations and discussions.}
\begin{center}
\subfigure[]{\includegraphics[scale=0.1]{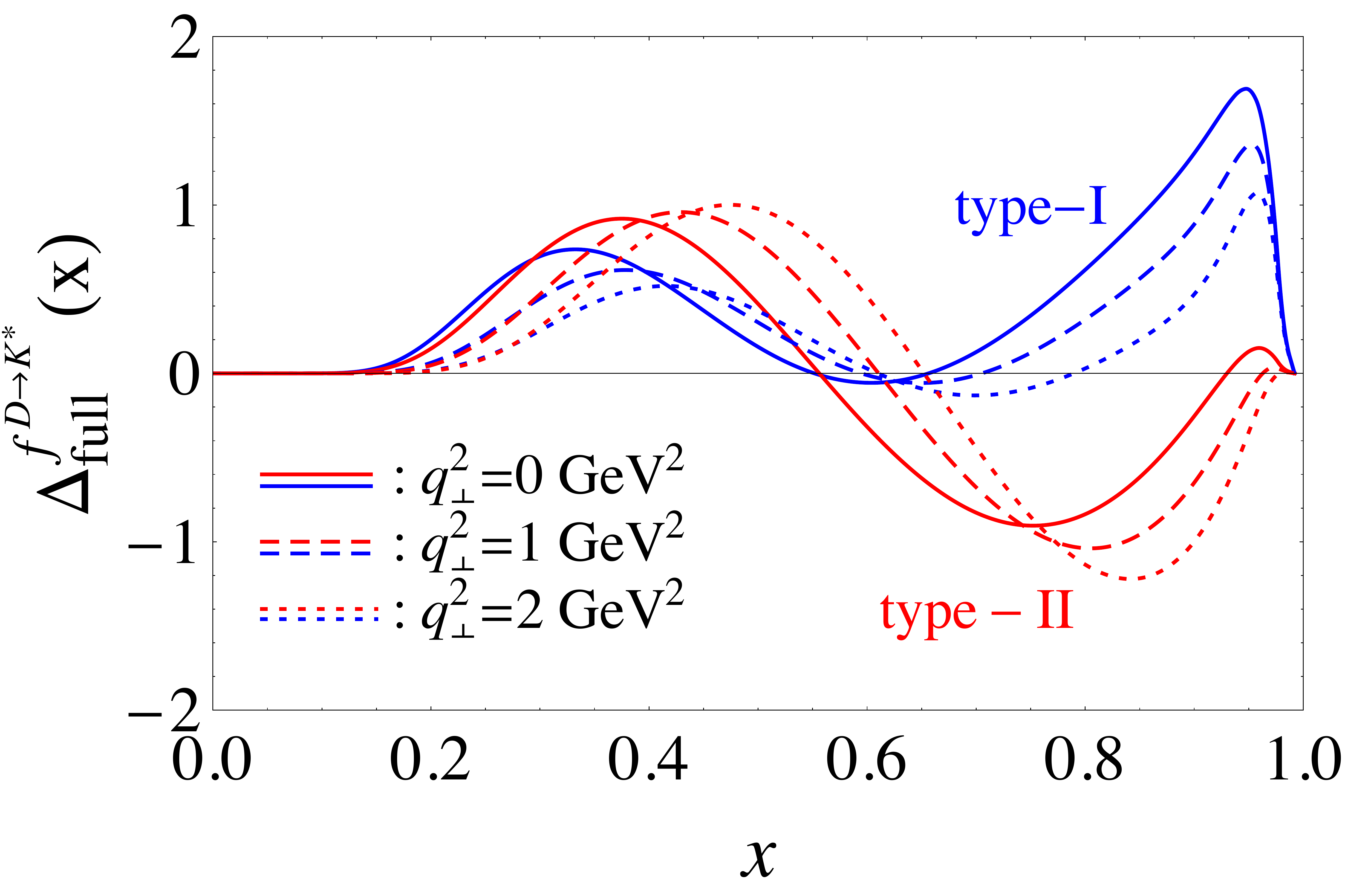}}\qquad\qquad
\subfigure[]{\includegraphics[scale=0.1]{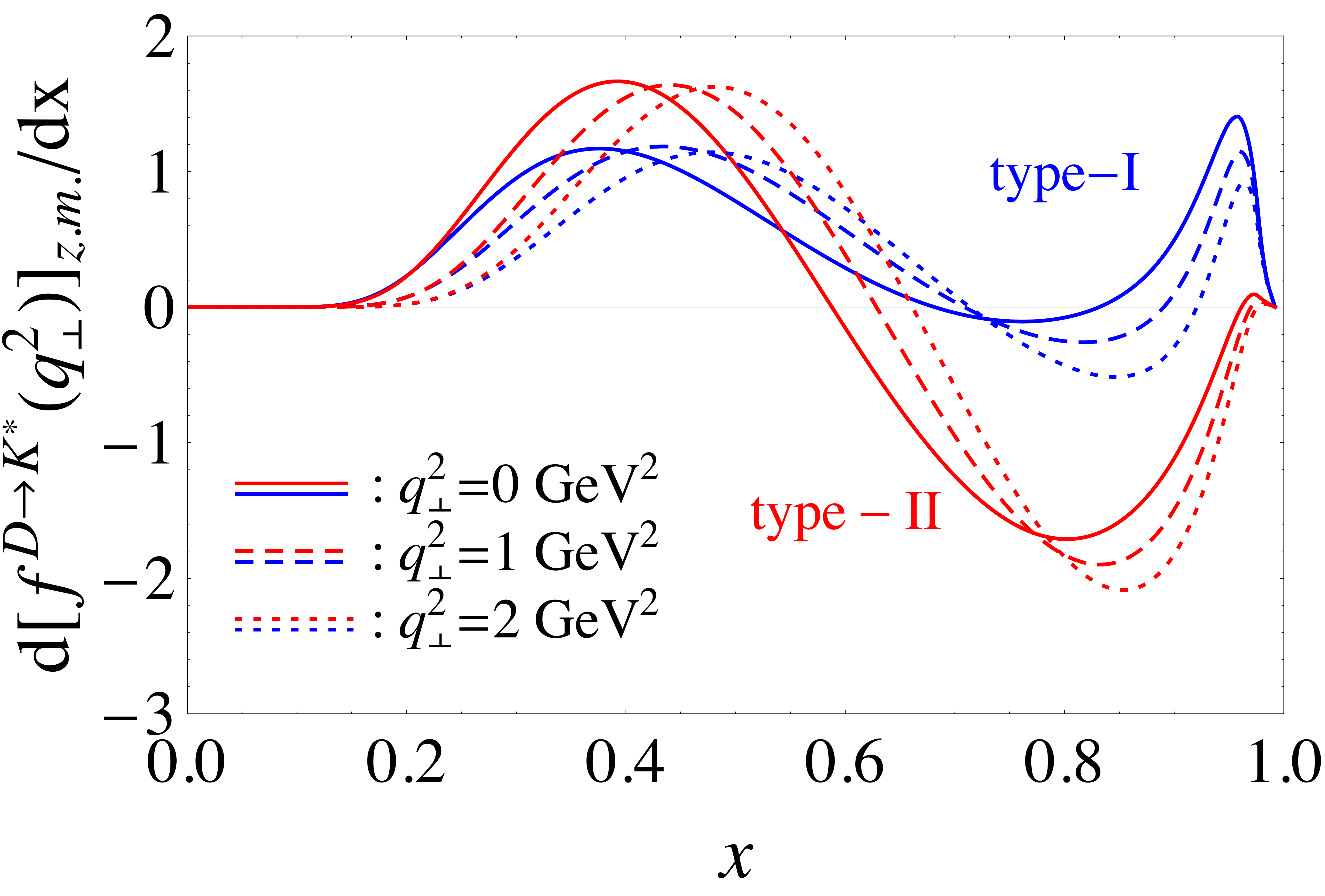}}\\
\subfigure[]{\includegraphics[scale=0.105]{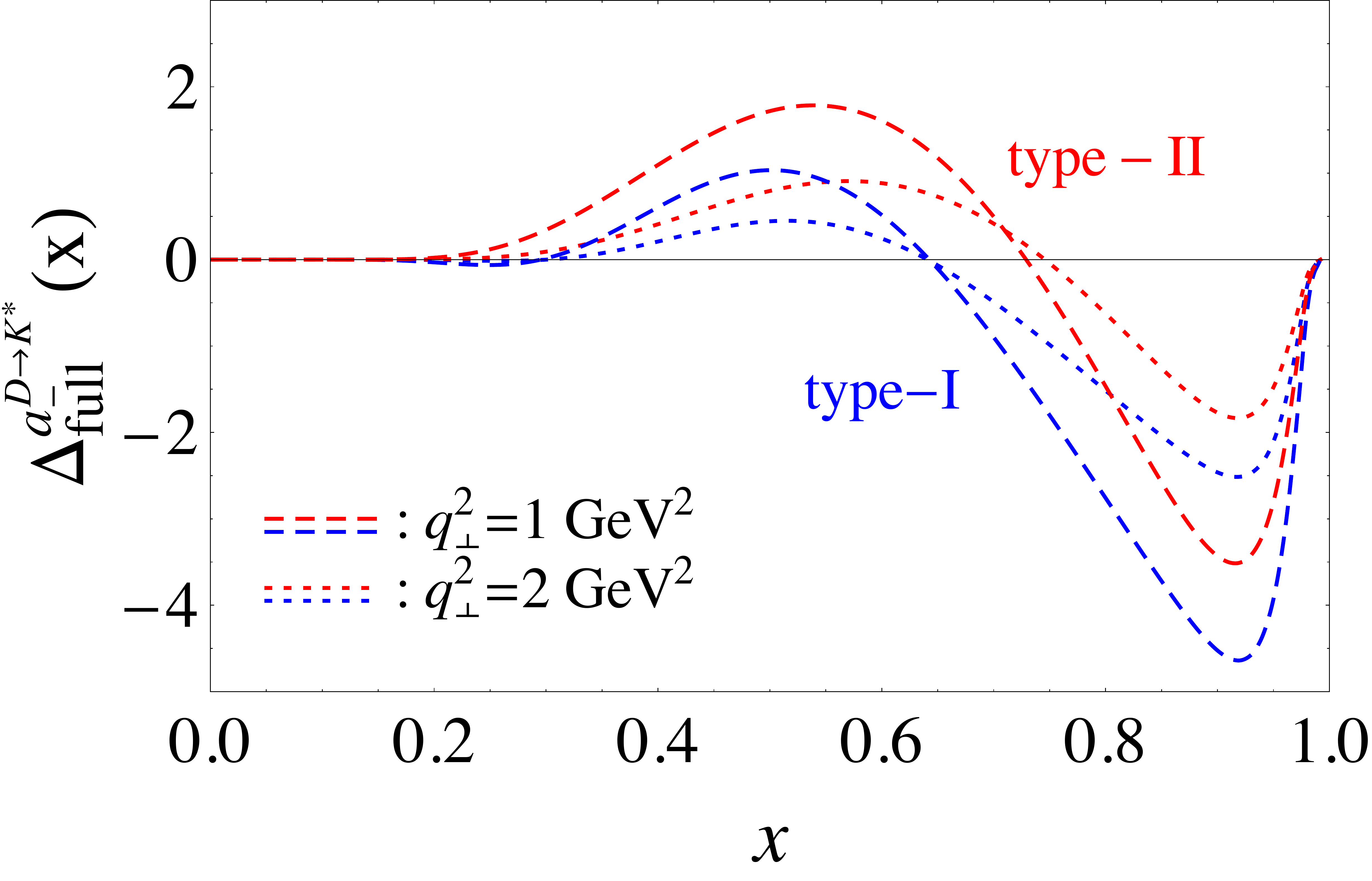}}\qquad\qquad
\subfigure[]{\includegraphics[scale=0.105]{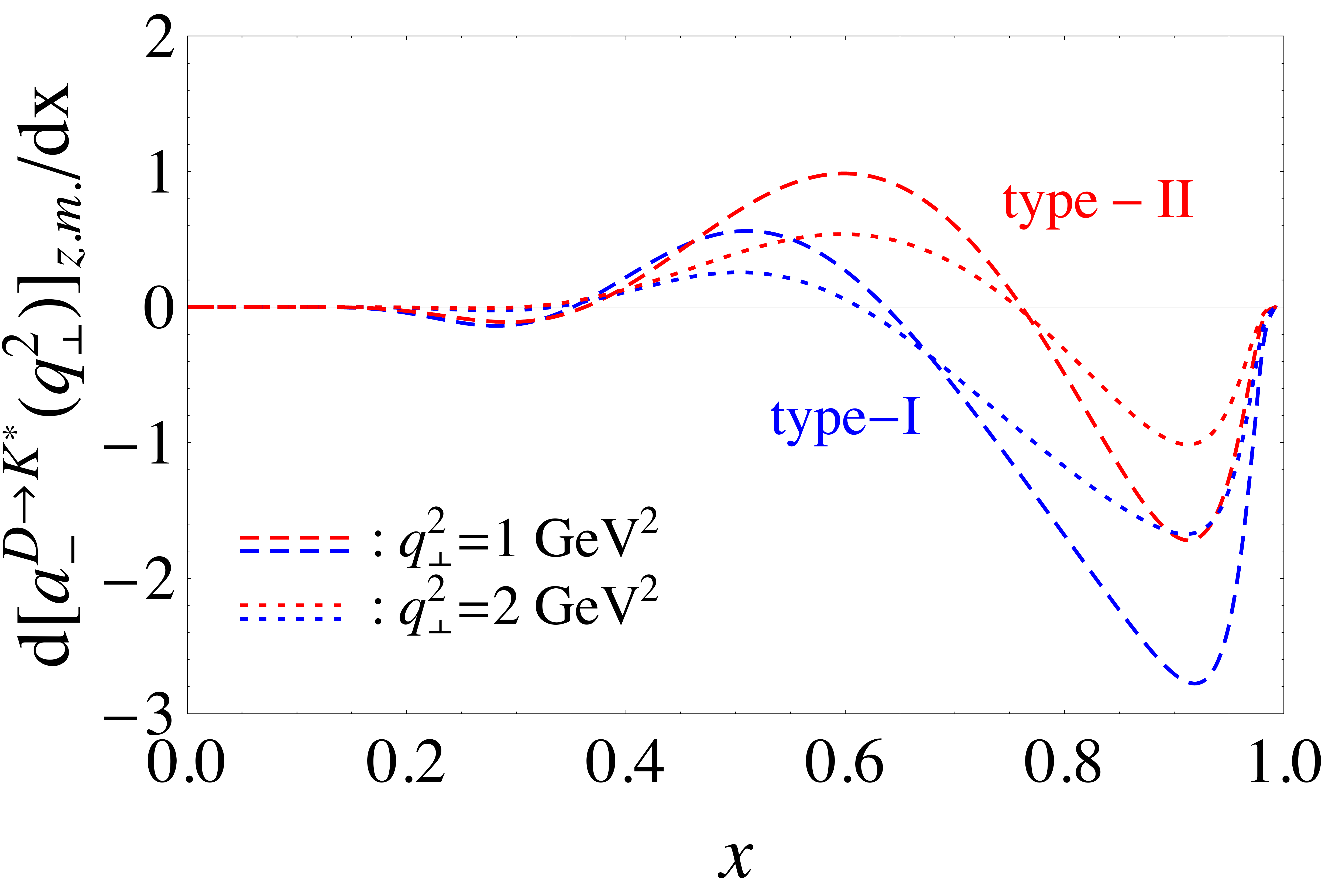}}
\end{center}
\label{fig:DK}
\end{figure}

Based on the theoretical results given above, we then present our numerical results and discussions. In the numerical analyses, we use the values of Gaussian parameter $\beta$ obtained by fitting to the data of zero-mode independent $f_P$~\cite{Chang:2018zjq}.  In order to clearly show the self-consistency of CLF QM, {\it i.e.} the difference between $[{\cal F}]_{\rm full}^{\lbd=0}$ and $[{\cal F}]_{\rm full}^{\lbd=\pm}$, we define
\begin{align}
\Delta^{\cal F}_{\rm full}(x) \equiv \frac{\d [{\cal F}]_{\rm full}^{\lbd=0}}{ \d x}-\frac{\d [{\cal F}]_{\rm full}^{\lbd=\pm}}{ \d x}\,,
\end{align}
which is equal to zero for  $g(q^2)$ and $a_+(q^2)$ and related to $B_1^{(2)}$ and  $B_3^{(3)}$ functions for  $f(q^2)$ and $a_-(q^2)$. In addition, for convenience of analyses and discussions, we take the $B\to D^*$ and $D\to K^*$ transitions  as examples.  Using the central values of $\beta$, we summarize our numerical results for the form factors at different ${\bf q}_{\bot}^2$ in Tables \ref{tab:1}, \ref{tab:2} and  \ref{tab:3}, and show the dependences of $\Delta^{f\,,a_-}_{\rm full}(x)$ and $\d[{f\,,a_-}]_{\rm z.m.}/\d x$ on $x$ in Figs.~\ref{fig:BD} and \ref{fig:DK}.   Based on these numerical results and the theoretical formulae given in the last subsection, we have the following discussions and findings:
\begin{itemize}
\item From  Eqs.~\eqref{eq:gfull} and \eqref{eq:apfull}, it can be found that the CLF results for $g(q^2)$ and $a_+(q^2)$ are independent of the $B$ function contributions and the choice of $\lbd$, which implies that such two form factors do not have the self-consistency problem, {\it i.e.}, 
\begin{align}
[g(q^2)]_{\rm full}^{\lbd=0}=[g(q^2)]_{\rm full}^{\lbd=\pm}\,,\qquad [a_+(q^2)]_{\rm full}^{\lbd=0}=[a_+(q^2)]_{\rm full}^{\lbd=\pm}\,. 
\end{align}
From  Eqs.~\eqref{eq:gval} and \eqref{eq:apval}, it can be found that they are also free from the zero-mode effect. Further comparing Eqs.~\eqref{eq:gSLF} with~\eqref{eq:gfull} for $g(q^2)$ and Eqs.~\eqref{eq:apSLF} with~\eqref{eq:apfull} for $a_+(q^2)$, respectively, we can conclude that 
\begin{align}
 [g(q^2)]_{\rm SLF}=[g(q^2)]_{\rm val.}=[g(q^2)]_{\rm full}\,,\qquad  [a_+(q^2)]_{\rm SLF}=[a_+(q^2)]_{\rm val.}=[a_+(q^2)]_{\rm full}
\end{align}
within both the type-I and the type-II scheme. The numerical examples for $g(q^2)$ and $a_+(q^2)$ are given in Table \ref{tab:1}.

\item The form factors $f(q^2)$ and $a_-(q^2)$ are zero-mode dependent, which is different from the case of $g(q^2)$ and $a_+(q^2)$; and moreover, their CLF results extracted via $\lbd=0$ mode~($[f(q^2)]_{\rm full}^{\lbd=0}$ and $[a_-(q^2)]_{\rm full}^{\lbd=0}$)  receive additional contributions associated with $B_1^{(2)}$ and  $B_3^{(3)}$ functions relative to the ones obtained via $\lbd=\pm$ mode~($[f(q^2)]_{\rm full}^{\lbd=\pm}$ and $[a_-(q^2]_{\rm full}^{\lbd=\pm}$), which can be found from Eqs.~\eqref{eq:ffull} and \eqref{eq:amfull}.  From  Figs.~\ref{fig:BD}(a,c) and \ref{fig:DK}(a,c), it can be clearly seen that the contributions of  $B_1^{(2)}$ and  $B_3^{(3)}$ functions, $\int_0^1\d x \Delta^{f\,,a_-}_{\rm full}(x)$, are nonzero in the traditional type-I correspondence scheme. It also can be found from the numerical examples given in Tables \ref{tab:2} and \ref{tab:3}~(the fourth and fifth columns), from which one can further find that the effects of these contributions are very significant at some $q^2$ points.  Therefore, the CLF results in  type-I  scheme suffer from the self-consistency problems, $[f(q^2)]_{\rm full}^{\lbd=0}\neq [f(q^2)]_{\rm full}^{\lbd=\pm}$ and $[a_-(q^2)]_{\rm full}^{\lbd=0}\neq [a_-(q^2)]_{\rm full}^{\lbd=\pm}$~(type-I). 

Interestingly, within the type-II correspondence scheme,  the positive $\Delta^{f\,,a_-}_{\rm full}(x)$ with small $x$ and the negative one with large $x$ can exactly cancel each other at each $q^2$ point, therefore $\int\d x\Delta^{f\,,a_-}_{\rm full}(x)=0$, which are clearly shown in Figs.~\ref{fig:BD}(a,c) and \ref{fig:DK}(a,c). As a result, we find that 
\begin{equation}\label{eq:selfcon}
[f(q^2)]_{\rm full}^{\lbd=0}\;\dot{=}\;[f(q^2)]_{\rm full}^{\lbd=\pm}\,,\quad [a_-(q^2)]_{\rm full}^{\lbd=0}\;\dot{=}\;[a_-(q^2)]_{\rm full}^{\lbd=\pm}\,, \qquad (\text{type-II})
\end{equation}
which can also be easily found by comparing the numerical  results given in the  fourth and the fifth columns of Tables \ref{tab:2} and \ref{tab:3}.  The Eq.~\eqref{eq:selfcon} implies that the self-consistency  problem can be resolved by employing the type-II correspondence scheme. 

\item Comparing Eqs.~\eqref{eq:fSLF} with  \eqref{eq:fval} and Eqs.~\eqref{eq:amSLF}  with \eqref{eq:amval}, respectively, we do not find significant relations between the SLF results and the valence contributions in the CLF approach for $f(q^2)$ and $a_-(q^2)$ within the type-I scheme. However, employing the type-II scheme and making some simplifications on these formulas, we find surprisingly that the SLF  and valence results  are exactly the same, 
 \begin{equation}\label{eq:VSLFval}
 [f(q^2)]_{\rm SLF}\,=\,[f(q^2)]_{\rm val.} \quad\text{and}\quad [a_-(q^2)]_{\rm SLF}\, = \,[a_-(q^2)]_{\rm val.}\,, \qquad(\text{type-II})
\end{equation}
which can also been clearly seen from the numerical results given in  Tables \ref{tab:2} and \ref{tab:3}.

In addition, we find that $[f(q^2)]_{\rm SLF, val.}$ and $[ a_-(q^2)]_{\rm SLF, val.}$ are divergent at $q^2=0$ in the type-I scheme, therefore their numerical results are labeled as ``---'' in Tables \ref{tab:2} and \ref{tab:3}. It is  possibly the reason for why the author of  Ref.~\cite{Jaus:1999zv} claims in the abstract that $a_-(q^2)$ is impossible to be calculated in the SLF QM, while we  note that the CLF results,  $[f(0)]_{\rm full}^{\lbd=0}$ and $[a_-(0)]_{\rm full}^{\lbd=0}$, also suffer from such problem in the type-I scheme.  Interestingly, this divergence problem does not exist in the type-II scheme, and their numerical results at $q^2=0$ satisfy the relations given by Eqs. \eqref{eq:selfcon}, \eqref{eq:VSLFval} and following Eq.~\eqref{eq:fullval}.

\item  From Figs.~\ref{fig:BD}(b,d) and \ref{fig:DK}(b,d),  it can be seen that  zero-mode presents sizable contributions to $f(q^2)$ and $a_-(q^2)$ within the traditional type-I correspondence scheme, i.e.,  $[f(q^2)]_{\rm z.m.}\neq 0$ and  $[a_-(q^2)]_{\rm z.m.}\neq 0$~(type-I); while, in the  type-II correspondence scheme, these contributions, although existing formally, vanish numerically, i.e.,  $[f(q^2)]_{\rm z.m.}\dot{=} 0$ and  $[a_-(q^2)]_{\rm z.m.}\dot{=} 0$~(type-II). Therefore, one can further find that 
 \begin{equation}\label{eq:fullval}
 [f(q^2)]_{\rm full}\,\dot{=}\,[f(q^2)]_{\rm val.} \quad\text{and}\quad  [a_-(q^2)]_{\rm full}\,\dot{=}\,[a_-(q^2)]_{\rm val.} \,, \qquad(\text{type-II})
\end{equation}
which can also be found from the numerical examples given in Tables \ref{tab:2} and \ref{tab:3}. 

\end{itemize}

Combining the findings given above, we can finally conclude that
 \begin{equation}\label{eq:SLFfullval}
 [{\cal F}(q^2)]_{\rm SLF}\,=\,[{\cal F}(q^2)]_{\rm val.}\,\dot{=}\,[{\cal F}(q^2)]_{\rm full} \,, \qquad(\text{type-II})
\end{equation}
where, $[{\cal F}(q^2)]_{\rm full}\equiv [{\cal F}(q^2)]_{\rm full}^{\lbd=\pm}\dot{=}[{\cal F}(q^2)]_{\rm full}^{\lbd=0}$~[Eq.~\eqref{eq:selfcon}] implies the self-consistency of the  CLF QM and holds only in the  type-II correspondence scheme; the symbol, ``$\dot{=}$'', should be replaced by ``$=$" for $g(q^2)$ and $a_+(q^2)$.


Besides of the self-consistency problem, the CLF QM with traditional type-I correspondence scheme  also has a problem of  covariance, which will be discussed in the follows. 
A peculiar property of the LF matrix element is its dependence on the light-like four vector $\w=(0,2,{\bf 0}_{\bot})$~\cite{Jaus:1999zv,Jaus:2002sv}, which can be explicitly revealed by the decomposition
\begin{align}
\hat{{\cal B}}=\text{physical part}+\text{$\w$-dependent part}\,,
\end{align}
where the ``physical part'' contains the physical contributions to the form factors,  while the ``$\w$-dependent part'' is unphysical and may violate the covariance of matrix element if it is nonzero. In the CLF QM, most of the $\w$-dependent contributions are eliminated by the zero-mode contributions, but there are still some residual $\w$-dependences which are associated with $B$ functions and independent of  zero-mode~\cite{Jaus:1999zv}. Therefore, a different mechanism is required to ``neutralize'' the effects of these residual $\w$-dependent  contributions. 

In order to clearly show the residual $\w$-dependence of matrix element of $P\to V$ transition, after integrating out the $k^-$ component and taking into account the zero-mode contributions,  we can decompose the trace term $\hat{S}_{\cal B}$ in the integrand of Eq.~\eqref{eq:Bclf2} for $\la V(p^{\prime\prime},\lambda) | \bar q_1^{\prime\prime}\r_\u\r^5 q_1^\prime|P(p^\prime)\ra$ as
\begin{align}\label{eq:S1}
\hat{S}_{P\to V}^{\mu}
= & 4\, \frac{P^\u \e^*\cdot \w+\w^\u \e^*\cdot P}{\w\cdot P}\, \bigg\{ 2(m_1'-m_2)  B_1^{(2)}\nonumber\\
 & -\frac{1}{D_{\rm V, con}''}\Big[ \big(M'^2+M''^2-q^2 + 2(m_1'-m_2)(m_1''+m_2)\big)B_1^{(2)} -2B_3^{(3)}\Big]   \bigg\}+\text{...} \,,
\end{align}
where ``$...$'' denotes the physical contribution resulting in the CLF results for  $f(q^2)$ and $a_{\pm}(q^2)$ without considering contributions related to $B$ functions. For the first term in Eq. \eqref{eq:S1}, in order to separate the potential  physical contribution from the unphysical one, we  shall use the identity~\cite{Jaus:2002sv}  
 \begin{align}\label{eq:dcomp}
P^\u  \frac{\e\cdot \w}{\w\cdot P}=&\e^\u-{\frac{q^\u}{q^2}}\left({\e\cdot q}-q\cdot P \frac{\w\cdot\e}{\w\cdot P}\right)-\frac{\w^\u}{\w\cdot P} \left[ \e\cdot P-\e\cdot q \frac{q\cdot P}{q^2}-\e\cdot\w \frac{P^2}{\w\cdot P}+\e\cdot\w\frac{(q\cdot P)^2}{q^2\w\cdot P} \right] \nonumber\\
&-\frac{i\lbd \,}{\,\w\cdot P}\frac{\e\cdot q}{q^2}\ve^{\u\a\b\v}\w_\a q_\b P_\v\,.
 \end{align}

For the case of $\lbd=0$, using Eq.~\eqref{eq:dcomp}, the pre-factor in  Eq.~\eqref{eq:S1} can be written as
  \begin{align}\label{eq:pf0}
 \frac{P^\u  \e^*\cdot \w +\w^\u  \e^*\cdot P}{\w\cdot P}\Big|_{\lbd=0}
 =&\e^{*\u}-{\frac{q^\u}{q^2}}{\e^*\cdot q}\,+q^\u  \frac{q\cdot P\, \w\cdot\e^*}{q^2 \,\w\cdot P}  
 +\w^\u \frac{2M''}{\w\cdot P}\,.
 \end{align}
 For the case of $\lbd=\pm$,  instead of using Eq.~\eqref{eq:dcomp}, we can directly write  the pre-factor in  Eq.~\eqref{eq:S1} as
   \begin{align}\label{eq:pfpm}
 \frac{P^\u  \e^*\cdot \w +\w^\u  \e^*\cdot P}{\w\cdot P} \Big|_{\lbd=\pm}
 =\w^\u \frac{\e^*\cdot P}{\w\cdot P}\,,
 \end{align}
 due to $\e^*_{\lbd=\pm}\cdot \w=0 $.  Based on these formulae, we have following remarks:
 \begin{itemize}
\item Comparing Eq.~\eqref{eq:pf0} with Eq.~\eqref{eq:pfpm}, it can be found that the first and the second term in  Eq.~\eqref{eq:pf0} give additional contributions associated with $B$ functions to $[f(q^2)]_{\rm full}^{\lbd=0}$ and $[a_-(q^2)]_{\rm full}^{\lbd=0}$, respectively,  which results in the self-consistency problems of CLF QM.

The last terms in Eq.~\eqref{eq:pf0} and Eq.~\eqref{eq:pfpm} are the residual $\w$-dependent parts, which give the contributions to the unphysical form factor and will violate the covariance if nonzero. Therefore, we can conclude that the self-consistency and the covariance problems of CLF QM are in fact have the same origin. 

\item In the traditional type-I correspondence scheme, the residual $\w$-dependent parts in Eq.~\eqref{eq:pf0} and Eq.~\eqref{eq:pfpm}  are nonzero, therefore, the covariance of matrix element is violated. While, these unphysical $\w$-dependent parts  vanish numerically  in the type-II scheme because they are proportional to $\int  \d x \Delta^{ f,a_-}_{\rm full}(x)=0$ (type-II). It implies that the covariance can be recovered by employing the  type-II scheme.

\item  It should be noted that, for the $\lbd=\pm$ mode, the decomposition into physical and unphysical contributions is ambiguous. Instead of  Eq.~\eqref{eq:pfpm} used in this paper, one can also decompose  $\frac{P^\u  \e^*\cdot \w +\w^\u  \e^*\cdot P}{\w\cdot P} \big|_{\lbd=\pm}$ in the same way as $\frac{P^\u  \e^*\cdot \w +\w^\u  \e^*\cdot P}{\w\cdot P} \big|_{\lbd=0}$ by using Eq.~\eqref{eq:dcomp}. At this time, $[f(q^2)]_{\rm full}^{\lbd=\pm}$ and $[a_-(q^2)]_{\rm full}^{\lbd=\pm}$ have the same forms as $[f(q^2)]_{\rm full}^{\lbd=0}$ and $[a_-(q^2)]_{\rm full}^{\lbd=0}$, respectively, thus the self-consistency problem vanishes, which however is at the expense of introducing some unphysical form factors, for instance, the one corresponding to non-vanishing $\ve^{\u\a\b\v}\w_\a q_\b P_\v$ in Eq.~\eqref {eq:dcomp}. Therefore, we can conclude that whether the self-consistency  appears is in fact determined by the way of  decomposition for the contribution of $B$ functions.  This ambiguous decomposition  become trivial only when the contributions of $B$ functions are zero, which is impossible in the type-I corresponding scheme but can be achieved in the type-II scheme.
\end{itemize}

\begin{table}[ht]
\footnotesize 
\begin{center}
\caption{\label{tab:pred} \small The updated results of form factors for some $P\to V$ transitions. }
\vspace{0.1cm}
\let\oldarraystretch=\arraystretch
\renewcommand*{\arraystretch}{1.1}
\setlength{\tabcolsep}{2.8pt}
\begin{tabular}{lccc|ccccccccc}
\hline\hline
     $F$ &$F(0)$ &a &b  &$F$ &$F(0)$ &a &b
                \\\hline
$V^{D\to\rho}$      &$1.05^{+0.01+0.07}_{-0.01-0.07}$&$1.20^{+0.02+0.10}_{-0.02-0.10}$&$0.23^{+0.02+0.01}_{-0.02-0.01}$
&$A_0^{D\to\rho}$      &$0.68^{+0.01+0.04}_{-0.01-0.04}$&$1.27^{+0.02+0.04}_{-0.02-0.05}$&$0.30^{+0.02+0.03}_{-0.02-0.03}$
\\
$A_1^{D\to\rho}$    &$0.67^{+0.01+0.05}_{-0.01-0.05}$&$0.57^{+0.01+0.06}_{-0.01-0.04}$&$-0.02^{+0.00+0.01}_{-0.01-0.01}$
&$A_2^{D\to\rho}$      &$0.49^{+0.00+0.04}_{-0.00-0.04}$&$0.71^{+0.02+0.08}_{-0.02-0.07}$&$0.15^{+0.01+0.01}_{-0.01-0.01}$
\\
$V^{D\to K^*}$      &$1.14^{+0.01+0.06}_{-0.01-0.07}$&$1.09^{+0.02+0.08}_{-0.02-0.07}$&$0.21^{+0.02+0.01}_{-0.02-0.01}$
&$A_0^{D\to K^*}$      &$0.76^{+0.01+0.04}_{-0.01-0.05}$&$1.14^{+0.02+0.04}_{-0.03-0.05}$&$0.26^{+0.02+0.03}_{-0.02-0.03}$
\\
$A_1^{D\to K^*}$    &$0.78^{+0.01+0.05}_{-0.01-0.06}$&$0.56^{+0.02+0.03}_{-0.02-0.03}$&$-0.02^{+0.01+0.01}_{-0.01-0.01}$
&$A_2^{D\to K^*}$      &$0.65^{+0.00+0.05}_{-0.00-0.06}$&$0.82^{+0.03+0.07}_{-0.03-0.07}$&$0.12^{+0.01+0.01}_{-0.01-0.01}$
\\\hline
$V^{D_s\to K^*}$      &$1.07^{+0.02+0.08}_{-0.02-0.08}$&$1.40^{+0.04+0.10}_{-0.04-0.09}$&$0.39^{+0.04+0.05}_{-0.04-0.05}$
&$A_0^{D_s\to K^*}$      &$0.59^{+0.01+0.04}_{-0.01-0.05}$&$1.50^{+0.04+0.10}_{-0.04-0.10}$&$0.51^{+0.04+0.02}_{-0.05-0.02}$
\\
$A_1^{D_s\to K^*}$    &$0.63^{+0.01+0.06}_{-0.01-0.06}$&$0.79^{+0.01+0.04}_{-0.01-0.03}$&$0.05^{+0.01+0.00}_{-0.02-0.00}$
&$A_2^{D_s\to K^*}$      &$0.53^{+0.00+0.06}_{-0.00-0.06}$&$0.98^{+0.05+0.09}_{-0.05-0.08}$&$0.24^{+0.02+0.01}_{-0.03-0.01}$
\\
$V^{D_s\to \phi}$      &$1.24^{+0.01+0.06}_{-0.01-0.06}$&$1.21^{+0.02+0.05}_{-0.02-0.05}$&$0.30^{+0.02+0.05}_{-0.02-0.05}$
&$A_0^{D_s\to \phi}$      &$0.71^{+0.00+0.04}_{-0.01-0.05}$&$1.29^{+0.02+0.06}_{-0.02-0.06}$&$0.38^{+0.02+0.01}_{-0.02-0.01}$
\\
$A_1^{D_s\to \phi}$    &$0.77^{+0.00+0.06}_{-0.00-0.07}$&$0.70^{+0.02+0.01}_{-0.02-0.01}$&$0.03^{+0.01+0.00}_{-0.01-0.00}$
&$A_2^{D_s\to \phi}$      &$0.66^{+0.00+0.06}_{-0.00-0.07}$&$0.92^{+0.02+0.06}_{-0.02-0.06}$&$0.18^{+0.01+0.01}_{-0.01-0.01}$
\\\hline
$V^{B\to \rho}$      &$0.35^{+0.01+0.06}_{-0.01-0.05}$&$1.70^{+0.02+0.20}_{-0.02-0.20}$&$0.84^{+0.04+0.10}_{-0.04-0.10}$
&$A_0^{B\to \rho}$      &$0.30^{+0.01+0.05}_{-0.01-0.05}$&$1.76^{+0.02+0.20}_{-0.02-0.20}$&$0.97^{+0.04+0.20}_{-0.05-0.17}$
\\
$A_1^{B\to \rho}$    &$0.27^{+0.01+0.05}_{-0.01-0.04}$&$0.85^{+0.02+0.08}_{-0.02-0.06}$&$0.12^{+0.01+0.03}_{-0.01-0.02}$
&$A_2^{B\to \rho}$      &$0.25^{+0.01+0.04}_{-0.01-0.04}$&$1.44^{+0.03+0.20}_{-0.03-0.13}$&$0.64^{+0.04+0.09}_{-0.04-0.07}$
\\
$V^{B\to K^*}$      &$0.40^{+0.01+0.07}_{-0.01-0.06}$&$1.65^{+0.03+0.20}_{-0.03-0.20}$&$0.80^{+0.05+0.10}_{-0.06-0.10}$
&$A_0^{B\to K^*}$      &$0.35^{+0.01+0.06}_{-0.01-0.06}$&$1.71^{+0.03+0.20}_{-0.03-0.20}$&$0.91^{+0.05+0.10}_{-0.06-0.10}$
\\
$A_1^{B\to K^*}$    &$0.32^{+0.01+0.06}_{-0.01-0.05}$&$0.87^{+0.03+0.08}_{-0.03-0.06}$&$0.12^{+0.02+0.03}_{-0.02-0.02}$
&$A_2^{B\to K^*}$      &$0.30^{+0.01+0.05}_{-0.01-0.05}$&$1.46^{+0.03+0.20}_{-0.04-0.20}$&$0.64^{+0.05+0.08}_{-0.05-0.06}$
\\
$V^{B\to D^*}$      &$0.78^{+0.01+0.09}_{-0.01-0.10}$&$1.26^{+0.01+0.10}_{-0.01-0.10}$&$0.37^{+0.01+0.04}_{-0.02-0.03}$
&$A_0^{B\to D^*}$      &$0.68^{+0.01+0.08}_{-0.01-0.08}$&$1.28^{+0.01+0.10}_{-0.01-0.10}$&$0.40^{+0.02+0.05}_{-0.02-0.03}$
\\
$A_1^{B\to D^*}$    &$0.66^{+0.01+0.08}_{-0.01-0.08}$&$0.66^{+0.01+0.05}_{-0.01-0.05}$&$0.00^{+0.01+0.01}_{-0.01-0.01}$
&$A_2^{B\to D^*}$      &$0.62^{+0.00+0.08}_{-0.00-0.08}$&$1.13^{+0.02+0.10}_{-0.02-0.09}$&$0.30^{+0.02+0.03}_{-0.02-0.02}$
\\\hline
$V^{B_s\to K^*}$      &$0.28^{+0.02+0.07}_{-0.02-0.06}$&$2.06^{+0.04+0.20}_{-0.05-0.20}$&$1.82^{+0.14+0.20}_{-0.15-0.10}$
&$A_0^{B_s\to K^*}$      &$0.22^{+0.01+0.06}_{-0.01-0.05}$&$2.14^{+0.04+0.20}_{-0.05-0.20}$&$2.05^{+0.15+0.20}_{-0.17-0.20}$
\\
$A_1^{B_s\to K^*}$    &$0.20^{+0.01+0.05}_{-0.01-0.05}$&$1.29^{+0.05+0.06}_{-0.05-0.04}$&$0.61^{+0.07+0.05}_{-0.08-0.03}$
&$A_2^{B_s\to K^*}$      &$0.19^{+0.01+0.05}_{-0.01-0.04}$&$1.80^{+0.05+0.20}_{-0.06-0.16}$&$1.45^{+0.12+0.10}_{-0.14-0.08}$
\\
$V^{B_s\to \phi}$      &$0.38^{+0.01+0.09}_{-0.01-0.08}$&$1.91^{+0.02+0.20}_{-0.02-0.13}$&$1.44^{+0.06+0.10}_{-0.07-0.10}$
&$A_0^{B_s\to \phi}$      &$0.30^{+0.01+0.07}_{-0.01-0.06}$&$1.98^{+0.02+0.20}_{-0.02-0.20}$&$1.61^{+0.07+0.20}_{-0.07-0.20}$
\\
$A_1^{B_s\to \phi}$    &$0.28^{+0.01+0.07}_{-0.01-0.06}$&$1.18^{+0.02+0.06}_{-0.03-0.05}$&$0.44^{+0.03+0.04}_{-0.03-0.03}$
&$A_2^{B_s\to \phi}$      &$0.26^{+0.01+0.07}_{-0.01-0.06}$&$1.69^{+0.03+0.20}_{-0.03-0.17}$&$1.16^{+0.06+0.08}_{-0.06-0.07}$
\\
$V^{B_s\to D_s^*}$      &$0.83^{+0.01+0.10}_{-0.01-0.10}$&$1.34^{+0.03+0.10}_{-0.03-0.08}$&$0.52^{+0.03+0.03}_{-0.03-0.02}$
&$A_0^{B_s\to D_s^*}$      &$0.68^{+0.01+0.09}_{-0.01-0.09}$&$1.37^{+0.03+0.10}_{-0.03-0.10}$&$0.57^{+0.03+0.03}_{-0.04-0.02}$
\\
$A_1^{B_s\to D_s^*}$    &$0.66^{+0.01+0.09}_{-0.01-0.10}$&$0.76^{+0.03+0.04}_{-0.03-0.02}$&$0.10^{+0.01+0.03}_{-0.01-0.03}$
&$A_2^{B_s\to D_s^*}$      &$0.59^{+0.00+0.08}_{-0.00-0.09}$&$1.17^{+0.04+0.08}_{-0.04-0.06}$&$0.41^{+0.03+0.01}_{-0.03-0.01}$
\\
\hline
$V^{B_c\to D^*}$      &$0.23^{+0.02+0.09}_{-0.02-0.07}$&$3.51^{+0.09+0.20}_{-0.10-0.13}$&$7.91^{+0.67+0.10}_{-0.70-0.10}$
&$A_0^{B_c\to D^*}$      &$0.13^{+0.01+0.05}_{-0.01-0.04}$&$3.64^{+0.09+0.20}_{-0.10-0.20}$&$8.79^{+0.72+0.30}_{-0.78-0.30}$
\\
$A_1^{B_c\to D^*}$    &$0.13^{+0.01+0.05}_{-0.01-0.04}$&$2.76^{+0.10+0.05}_{-0.11-0.01}$&$4.56^{+0.47+0.30}_{-0.50-0.30}$
&$A_2^{B_c\to D^*}$      &$0.13^{+0.01+0.05}_{-0.01-0.04}$&$3.04^{+0.11+0.10}_{-0.11-0.09}$&$6.13^{+0.59+0.10}_{-0.61-0.10}$
\\
$V^{B_c\to D_s^*}$      &$0.42^{+0.03+0.10}_{-0.03-0.10}$&$2.99^{+0.10+0.20}_{-0.10-0.17}$&$4.84^{+0.46+0.10}_{-0.51-0.10}$
&$A_0^{B_c\to D_s^*}$      &$0.24^{+0.02+0.07}_{-0.02-0.06}$&$3.11^{+0.10+0.20}_{-0.10-0.20}$&$5.41^{+0.50+0.20}_{-0.58-0.20}$
\\
$A_1^{B_c\to D_s^*}$    &$0.24^{+0.02+0.08}_{-0.02-0.06}$&$2.23^{+0.11+0.06}_{-0.11-0.05}$&$2.53^{+0.30+0.20}_{-0.36-0.14}$
&$A_2^{B_c\to D_s^*}$      &$0.21^{+0.01+0.07}_{-0.01-0.06}$&$2.50^{+0.12+0.10}_{-0.12-0.10}$&$3.55^{+0.40+0.04}_{-0.47-0.04}$
\\
$V^{B_c\to J/\psi}$      &$0.90^{+0.01+0.20}_{-0.01-0.20}$&$2.25^{+0.03+0.04}_{-0.03-0.05}$&$2.08^{+0.08+0.06}_{-0.09-0.04}$
&$A_0^{B_c\to J/\psi}$      &$0.57^{+0.01+0.10}_{-0.01-0.10}$&$2.34^{+0.03+0.05}_{-0.03-0.06}$&$2.31^{+0.09+0.05}_{-0.09-0.05}$
\\
$A_1^{B_c\to J/\psi}$    &$0.57^{+0.01+0.10}_{-0.01-0.10}$&$1.16^{+0.03+0.20}_{-0.03-0.13}$&$1.05^{+0.05+0.10}_{-0.06-0.10}$
&$A_2^{B_c\to J/\psi}$      &$0.52^{+0.01+0.10}_{-0.01-0.10}$&$1.97^{+0.03+0.05}_{-0.03-0.06}$&$1.62^{+0.07+0.05}_{-0.07-0.02}$
\\\hline\hline
\end{tabular}
\end{center}
\end{table}

Therefore, we can conclude that the problems of self-consistency and covariance in the CLF QM can be resolved simultaneously by taking type-II correspondence scheme. Finally,  using the values of input parameters summarized in the appendix and employing the type-II scheme, we present our updated numerical results of BSW form factors for some $P\to V$ transitions in Table \ref{tab:pred}, where the two uncertainties are caused by parameters $\beta$ and quark masses, respectively. Some form factors have also been evaluated by other approaches, for instance, Lattice QCD~\cite{DelDebbio:1997ite,Flynn:1995dc}, light-cone sum rules~\cite{Ball:2004rg,Khodjamirian:2006st,Straub:2015ica,Cheng:2017smj} and perturbative QCD~\cite{Lu:2002ny,Wang:2007an,Wen-Fei:2013uea,Rui:2014tpa,Wang:2014yia,Hua:2018kho,Fan:2015kna,Liu:2018kuo}. Through comparison of these previous results with ours listed in Table \ref{tab:pred}, it is found that they are generally consistent with each other within theoretical uncertainties.

\section{Summary}
In this paper, we have investigated the self-consistency and Lorentz covariance of the CLF QM via the matrix elements and  relevant form factors of $P\to V$ transition, which provide much stricter tests on the CLF QM and are much more complicated than the case of decay constants studied in the previous works~\cite{Choi:2013mda,Chang:2018zjq}. Two types of correspondence schemes between  the  manifest covariant BS  approach and the LF QM are studied in detail. The main difference between these two schemes resides in whether the replacement $M\to M_0$ is applied only in the vertex operator or in each term in the integrand. Meanwhile, the results in the SLF QM are also presented for comparison. Our main findings are summarized as follows: 
\begin{itemize}
\item  The form factors $g(q^2)$ and $a_{+}(q^2)$ are independent of the spurious contributions associated with $B$ functions, therefore their CLF results obtained via $\lbd=0$ and $\pm$ polarization states of vector meson are consistent with each other, $[g(q^2)]_{\rm full}^{\lbd=0}=[g(q^2)]_{\rm full}^{\lbd=\pm}$ and $ [a_+(q^2)]_{\rm full}^{\lbd=0}=[a_+(q^2)]_{\rm full}^{\lbd=\pm}$, within both type-I and -II correspondence schemes.  Moreover, they are also free from the zero-mode contributions, therefore their valence contributions and full results  in the CLF  QM are the same. Besides, they are also equal to the SLF results. These relations can be summarized by Eq.~\eqref{eq:SLFfullval} in both type-I and -II schemes.

\item In the CLF QM, the form factors $a_{-}(q^2)$ and $f(q^2)$ receive the  contributions related to $B_1^{(2)}$ and $B_3^{(3)}$ functions, and these contributions obtained via $\lbd=0$ and $\pm$ states within the type-I scheme are different with each other, therefore, the CLF results for $a_{-}(q^2)$ and $f(q^2)$ in the  type-I scheme suffer from the problem of  self-consistency. This problem can be resolved by employing the   type-II correspondence because  the contributions associated with  $B$ functions vanish numerically after taking $M\to M_0$. 

\item The form factors $a_{-}(q^2)$ and $f(q^2)$ receive the zero-mode contributions. These contributions exist formally but vanish numerically in the  type-II  scheme, which results in $[f(q^2)]_{\rm full}\,\dot{=}\,[f(q^2)]_{\rm val.}$ and $[a_-(q^2)]_{\rm full}\,\dot{=}\,[a_-(q^2)]_{\rm val.}$. Further considering $[f(q^2)]_{\rm SLF}\,=\,[f(q^2)]_{\rm val.}$ and $[a_-(q^2)]_{\rm SLF}\,{=}\,[a_-(q^2)]_{\rm val.}$~(type-II), we can conclude that  the relation given by Eq.~\eqref{eq:SLFfullval}   holds still for the form factos $a_{-}(q^2)$ and $f(q^2)$ in the type-II scheme, but is violated in the type-I scheme.

\item The manifest covariance of the CLF result for $\la V| \bar q\r_\u\r^5 q|P\ra$ is violated within the traditional type-I correspondence scheme, but remarkably, can be recovered by employing the type-II correspondence. We further show that the self-consistency and covariance problems of CLF QM have the same origin; in addition, whether the self-consistency problem exists  is in fact determined by the way of decomposition for the contribution of B functions. This ambiguous decomposition become trivial in the  type-II correspondence scheme. 

\end{itemize}

Above findings confirm further  the conclusion obtained via the decay constants of vector and  axial-vector  mesons in the previous works~\cite{Choi:2013mda,Chang:2018zjq}.

\begin{appendix}

\section*{Appendix: Input parameters }
\begin{table}[h]
\begin{center}
\caption{\label{tab:input} \small The values of Gaussian parameters $\beta$ (in unit of MeV), where $q=u$ and $d$.}
\vspace{0.2cm}
\let\oldarraystretch=\arraystretch
\renewcommand*{\arraystretch}{1.1}
\setlength{\tabcolsep}{8.8pt}
\begin{tabular}{lcccccccccc}
\hline\hline
  &$\beta_{q\bar{q}}$    &$\beta_{s\bar{q}}$   &$\beta_{s\bar{s}}$
  &$\beta_{c\bar{q}}$    &$\beta_{c\bar{s}}$ \\
  \hline
  P meson
  &$314.1\pm0.5$ &$350.7\pm1.6$ &$377.7\pm1.4$ &$461.2\pm11.1$ &$543.2\pm9.5$\\\hline
  V meson
  &$312.4\pm5.8$ &$314.2\pm9.6$ &$350.5\pm5.0$ &$412.0\pm12.0$ &$514.1\pm18.5$ \\\hline
  &$\beta_{c\bar{c}}$    &$\beta_{b\bar{q}}$   &$\beta_{b\bar{s}}$   &$\beta_{b\bar{c}}$   &$\beta_{b\bar{b}}$    \\\hline
  P meson
  &$753.3\pm14.0$ &$540.7\pm9.6$
  &$601.9\pm7.4$ &$933.9\pm11.1$
  &$1382.4\pm50.0$
  \\\hline
  V meson
  &$684.4\pm6.7$ &$504.4\pm14.2$
  &$556.4\pm10.1$ &$863.4\pm32.8$ &$1370.1\pm11.2$
\\
\hline\hline
\end{tabular}
\end{center}
\end{table}

The constituent quark masses and  Gaussian parameters $\beta$ are essential inputs for calculating the form factors of $P\to V$ transition. For the former, their values have been suggested in many previous works, for instance, Refs.~\cite{Verma:2011yw,Jaus:2002sv,Cheng:2003sm,Geng:2016pyr}; in addition, they can also be obtained from the variational principle~\cite{Choi:2007yu,Choi:2007se,Choi:2015ywa}; but the errors are not evaluated in these previous works. In this work, we take~\cite{Verma:2011yw}
\begin{equation}\label{eq:qmass}
m_{u(d),s,c,b}=(0.25,0.45,1.40,4.64)\,{\rm GeV}\,,
\end{equation}
as default inputs, and  assign $10\%$ uncertainties to them which can cover roughly most of the values suggested in Refs.~\cite{Verma:2011yw,Jaus:2002sv,Cheng:2003sm,Geng:2016pyr,Choi:2007yu,Choi:2007se,Choi:2015ywa}.  For the Gaussian parameters $\beta$, we summarize their values in Table~\ref{tab:input}, which are obtained by fitting to the data of decay constants of $P$ and $V$ mesons~\cite{Chang:2018zjq} with the default values of quark masses, Eq.~\eqref{eq:qmass}, as inputs. In the fit, the theoretical formulas for the decay constants in the CLF QM with type-II correspondence scheme given in Ref.~\cite{Chang:2018zjq}  are used.  These values are used in our updated predictions for the BSW form factors, whose momentum dependence can be parameterized and reproduced via the three parameter form
\begin{align}
{\cal F}(q^2)=\frac{{\cal F}(0)}{1-a(q^2/M'^2)+b(q^2/M'^2)^2}\,.
\end{align}
Our updated numerical results of ${\cal F}(0)$, $a$ and $b$ for some $P\to V$  transitions are summarized in Table~\ref{tab:pred}.

\end{appendix}

\section*{Acknowledgements}
This work is supported by the National Natural Science Foundation of China (Grant Nos. 11875122 and 11475055) and the Program for Innovative Research Team in University of Henan Province (Grant No.19IRTSTHN018).

\end{document}